\documentstyle [a4p,epsf,12pt] {article}
\catcode`@=11 
\def \gsim{\mathrel{\mathpalette\@versim>}}
\def \lsim{\mathrel{\mathpalette\@versim<}}
\def \@versim#1#2{\lower0.4ex\vbox{\baselineskip\z@skip\lineskip\z@skip
     \lineskiplimit\z@\ialign{$\m@th#1\hfil##\hfil$%
     \crcr#2\crcr\sim\crcr}}}
\catcode`@=12 
\def\mumu{\mathrm{\mu^+ \mu^-}}

\def\Zzero{\mathrm{{{\rm Z}^0}}}

\def\ra{\rightarrow}
\def\Bp{\mathrm{B^+}}
\def\B0{\mathrm{B^0}}

\def\Bq{\mathrm{B_q^0}}
\def\Bqb{\mathrm{\bar B_q^0}}
\def\Bd{\mathrm{B_d^0}}

\def\taud{\tau_{\mathrm d}}
\def\Bs{\mathrm{B_s^0}}

\def\BB{\mathrm{B - \bar B}}

\def\ccbar{\mathrm{c\overline{c}}}
\def\bbbar{\mathrm{b\overline{b}}}
\def\qqbar{\mathrm{q\overline{q}}}
\def\uubar{\mathrm{u\overline{u}}}
\def\ddbar{\mathrm{d\overline{d}}}
\def\ssbar{\mathrm{s\overline{s}}}
\def\arr{\rightarrow}
\def\Z{\mathrm Z^0}
\def\Zbb{\mathrm{Z \arr \bbbar}}
\def\Zcc{\mathrm{Z \arr \ccbar}}

\def\Zuu{\mathrm{Z \arr \uubar}}
\def\dmd{\Delta m_{\mathrm{d}}}
\def\dms{\Delta m_{\mathrm{s}}}
\def\dmq{\Delta m_{\mathrm{q}}}
\def\Lb{\Lambda_{\mathrm{b}}}
\def\bcas{\mathrm{b \arr c \arr \ell}}
\def\bcb{\mathrm{b \arr \bar{c} \arr \ell}} 
\def\btau{\mathrm{b \arr \tau \arr \ell}}

\def\etal{{\sl et al.}}
\def\nn{\alpha_{\mathrm{kin}}}
\def\reps{{\mathrm{Re}}\,\epsilon_{\mathrm B}}
\def\idel{{\mathrm{Im}}\,\delta_{\mathrm B}}
\def\repsbs{{\mathrm{Re}}\,\epsilon_{\Bs}}
\def\idelbs{{\mathrm{Im}}\,\delta_{\Bs}}

\begin{document}           
\begin{titlepage}
\begin{center}
  {\large   EUROPEAN LABORATORY FOR PARTICLE PHYSICS }
\end{center}
\bigskip
\begin{tabbing}
\` CERN-PPE/97-036 \\
\` 2 April 1997 \\
\end{tabbing}
\vspace{1 mm}
\begin{center}{\LARGE\bf
A Study of B Meson Oscillations
}\end{center}
\begin{center}{\LARGE\bf
\mbox{\boldmath Using Hadronic $\Z$ Decays}
}\end{center}
\begin{center}{\LARGE\bf
Containing Leptons
}\end{center}
\vspace{10 mm}
\begin{center}{\LARGE
The OPAL Collaboration
}\end{center}
\vspace{6 mm}
\begin{abstract}
\vspace{5 mm}
An event sample enriched in semileptonic decays of b hadrons
is selected 
using an inclusive lepton selection
from approximately 3.0 million hadronic $\Z$ decays
collected with the OPAL detector.
This sample is used to investigate B meson oscillations
by reconstructing a proper decay time for the parent of each lepton,
using a jet charge method to estimate the production flavour of this
parent, and using the lepton charge to tag the decay flavour.
We measure the mass difference between the two $\Bd$ mass eigenstates
\[
\dmd = 0.444 \pm 0.029~^{+0.020}_{-0.017}~\mathrm{ps}^{-1} \; .
\]
For the $\Bs$ system,
we find $\dms>3.1$~ps$^{-1}$ at the 95\% confidence level.
This limit varies only a little if alternative limit setting
approaches are adopted.
Regions at higher $\dms$ values are also excluded with
some methods for setting the limit.

By studying the charge symmetry of the $\Bd$ mixing structure,
we are able to constrain possible CP and CPT violating
effects.
We measure the CP violation parameter
\[ \reps = -0.006 \pm 0.010 \pm 0.006
\]
and the indirect CPT violating parameter
\[ \idel = -0.020 \pm 0.016 \pm 0.006 \; .
\]
If we invoke CPT symmetry, then we obtain
\[
\reps = 0.002 \pm 0.007 \pm 0.003 \; .
\]
\parskip 0.5cm
\end{abstract}
\vspace{10mm}
\begin{center}
(Submitted to Zeitschrift f\"{u}r Physik C)
\end{center}
\end{titlepage}
\begin{center}{\Large        The OPAL Collaboration
}\end{center}\bigskip
\begin{center}{
K.\thinspace Ackerstaff$^{  8}$,
G.\thinspace Alexander$^{ 23}$,
J.\thinspace Allison$^{ 16}$,
N.\thinspace Altekamp$^{  5}$,
K.J.\thinspace Anderson$^{  9}$,
S.\thinspace Anderson$^{ 12}$,
S.\thinspace Arcelli$^{  2}$,
S.\thinspace Asai$^{ 24}$,
D.\thinspace Axen$^{ 29}$,
G.\thinspace Azuelos$^{ 18,  a}$
% ...
}\end{center}\bigskip
\bigskip
$^{  1}$School of Physics and Space Research, University of Birmingham,
Birmingham B15 2TT, UK
\newline
$^{  2}$Dipartimento di Fisica dell' Universit\`a di Bologna and INFN,
I-40126 Bologna, Italy
\newline
% ...
$^{  5}$Cavendish Laboratory, Cambridge CB3 0HE, UK
\newline
% ...
$^{  8}$CERN, European Organisation for Particle Physics,
CH-1211 Geneva 23, Switzerland
\newline
$^{  9}$Enrico Fermi Institute and Department of Physics,
University of Chicago, Chicago IL 60637, USA
\newline
% ...
$^{ 12}$Indiana University, Department of Physics,
Swain Hall West 117, Bloomington IN 47405, USA
\newline
% ...
$^{ 16}$Department of Physics, Schuster Laboratory, The University,
Manchester M13 9PL, UK
\newline
% ...
$^{ 18}$Laboratoire de Physique Nucl\'eaire, Universit\'e de Montr\'eal,
Montr\'eal, Quebec H3C 3J7, Canada
\newline
% ...
$^{ 23}$Department of Physics and Astronomy, Tel Aviv University,
Tel Aviv 69978, Israel
\newline
$^{ 24}$International Centre for Elementary Particle Physics and
Department of Physics, University of Tokyo, Tokyo 113, and
Kobe University, Kobe 657, Japan
\newline
% ...
$^{ 29}$University of British Columbia, Department of Physics,
Vancouver BC V6T 1Z1, Canada
\newline
% ...
$^{  a}$ and at TRIUMF, Vancouver, Canada V6T 2A3
\newline
% ...
\newpage
%
\section[intro]{Introduction}
The phenomenon of $\BB$ mixing is now well established.
Particle-antiparticle oscillations arise when the weak eigenstates,
$\mathrm{|B^0}\rangle$\footnote{
We use $\B0$ to refer to either of the neutral B mesons, $\Bd$ and $\Bs$.
}
 and $\mathrm{|\bar B^0}\rangle$,
differ from the mass eigenstates, 
$\mathrm{|B_1} \rangle$ and $\mathrm{|B_2} \rangle$,
which can be described as follows:
\begin{eqnarray}
 \mathrm{|B_1} \rangle & = &\frac{
   (1+\epsilon +\delta)\mathrm{|B^0}\rangle
 + (1-\epsilon -\delta)\mathrm{|\bar B^0}\rangle }
 {\sqrt{2(1+|\epsilon+\delta|^2)}} \nonumber \\
 \mathrm{|B_2} \rangle & = &\frac{
  (1+\epsilon -\delta)\mathrm{|B^0}\rangle  
- (1-\epsilon +\delta)\mathrm{|\bar B^0}\rangle }
 {\sqrt{2(1+|\epsilon-\delta|^2)}} \; ,
\end{eqnarray}
where $\epsilon$ and $\delta$ are complex, and parametrise
indirect CP and CPT violation \cite{potting}.
Note that $\epsilon$ allows for CP and T violation while respecting CPT symmetry,
and $\delta$ allows for CP and CPT violation, but respects T symmetry.
This formalism applies to both the $\Bd$ and $\Bs$ systems,
with separate values for $\epsilon$ and $\delta$ in each system.
The mass difference, $\dmq$, between the two mass eigenstates
for the $\Bq$-$\Bqb$ system determines the frequency of oscillation,
where $\Bq$ stands for either $\Bd$ or $\Bs$.
In the Standard Model, transitions between $\Bq$ and $\Bqb$ mesons
arise dominantly via box diagrams involving virtual top quarks.
Predictions for the mass differences can be made, and depend
on the top mass and the CKM elements $V_{\mathrm{tq}}$.
For the $\Bd$ system,
\begin{equation} \dmd \propto f^2_{\Bd} B_{\Bd} 
  m_{\mathrm{t}}^2 F(m_{\mathrm{t}}^2)
  |V_{\mathrm{td}} V^*_{\mathrm{tb}}|^2 \end{equation}
where the first two factors are the meson decay 
constant and QCD bag model vacuum insertion parameter, 
respectively.
These factors are obtained from 
lattice QCD calculations and QCD sum rules,
but there is an uncertainty of the order of 50\% on 
this product~\cite{ali}.
The next two factors are 
the top quark mass, $m_{\mathrm t}$, and a known quadratic
function of $m_{\mathrm t}$.
The last factor is a product of
CKM matrix elements.
At this point, an accurate measurement of $\dmd$ alone would not lead
to an accurate result for $V_{\mathrm{td}}$ because of 
the uncertainty in the decay and bag constants.
However, if both $\dmd$ and $\dms$ could be measured 
then information on the CKM matrix could be extracted via the ratio
\begin{equation} \frac{\dms}{\dmd} = \frac{m_{\Bs}}{m_{\Bd}} \cdot 
 \frac{|V_{\mathrm{ts}}|^2}{|V_{\mathrm{td}}|^2}
   \cdot \frac{f^2_{\Bs} B_{\Bs}}{f^2_{\Bd} B_{\Bd}} \; ,
\end{equation}
where $m_{\Bs}$ and $m_{\Bd}$ are the $\Bs$ and $\Bd$ masses,
as the ratio of decay constants for $\Bd$ and $\Bs$ mesons is much
better known than the absolute values \cite{ali,nir}.

CP violation has so far been observed
only in the $\mathrm K^0$ system, but it is also expected to
occur in the $\B0$ system.
Predictions for $\reps$ are of the order of $10^{-3}$ in the
Standard Model for the $\Bd$ system~\cite{CP}, and up
to an order of magnitude larger in the superweak model~\cite{sweak}.
The analogous parameter for the $\Bs$ system,
$\repsbs$, is expected to be smaller by at least
an order of magnitude, because 
${\mathrm{Im}}\, V_{\mathrm{ts}} \approx 
\lambda \cdot {\mathrm{Im}}\, V_{\mathrm{td}}$,
where $\lambda$ comes from the Wolfenstein parametrisation
of the CKM matrix~\cite{wolf}.
CPT violation may occur in certain
string description models of fundamental particles~\cite{potting2}.
Although CPT conservation has
been tested with high precision in the
$\mathrm K^0$ system~\cite{cplear}, CPT violating effects could be larger
in the B system~\cite{potting,alrick}.
 
The study presented in this paper uses inclusive lepton events,
where the identified lepton serves both to select $\Zbb$ events
and to determine the flavour of the parent b hadron at decay.
The production flavour of this b hadron is determined using a 
jet charge technique.
A vertex finding algorithm is used to estimate the decay length 
of the b hadron, and a separate algorithm is used to estimate the
energy of the hadron, allowing the decay proper time to be calculated.
The study of $\dmd$ and $\dms$ is performed neglecting possible CP
or CPT violation.
Results on $\reps$ and $\idel$ for the $\Bd$ system are then
obtained, assuming a negligible CP violating contribution  
from $\Bs$ decays.

\section{The OPAL Detector}
 
 
The OPAL detector has been described elsewhere~\cite{OPAL,opalsi}.
Tracking of charged particles is performed by a central detector,
consisting of a silicon microvertex detector,
a vertex chamber, a jet chamber
and $z$-chambers.\footnote{
The coordinate system is defined with
positive $z$ along the $\mathrm{e}^-$
beam direction, $\theta$ and
$\phi$ being the polar and azimuthal angles.
The origin is taken to
be the centre of the detector.}
The central detector is
positioned inside a
solenoid, which provides a uniform magnetic
field of 0.435 T.
The silicon microvertex detector consists of two layers of
silicon strip detectors;
the inner layer covers a polar angle range of
$| \cos \theta | < 0.83$ and
the outer layer covers $| \cos \theta |< 0.77$.
This detector provided both $\phi$- and $z$-coordinates for data taken in 
1993 and 1994, but $\phi$-coordinates only for 1991 and 1992.
Only $\phi$-coordinate information was used in this analysis.
The vertex chamber is a precision drift chamber
which covers the range $|\cos \theta | < 0.95$.
The jet chamber is
a large-volume drift chamber, 4.0~m long and 3.7~m in diameter,
providing both tracking and ionisation energy loss (d$E$/d$x$) information.
The $z$-chambers
measure the $z$-coordinate
of tracks as they leave the jet chamber in the range
$|\cos \theta | < 0.72$.
The coil is surrounded by a
time-of-flight counter array and
a barrel lead-glass electromagnetic calorimeter with a presampler.
Including the endcap electromagnetic calorimeters,
the lead-glass blocks
cover the range $| \cos \theta | < 0.98$.
The magnet return yoke 
is instrumented with streamer tubes
and serves as a hadron calorimeter.
Outside the hadron calorimeter are muon chambers, which
cover 93\% of the full solid angle.

\section{Event simulation}
 
Monte Carlo events are used to predict the relative abundances and
decay time distributions for lepton candidates from various
physics processes.
The JETSET~7.4 Monte Carlo program~\cite{jetset}
with parameters tuned to OPAL data~\cite{jset2}
was used to generate $\Zzero\ra\qqbar$ events
which were subsequently processed 
by the detector simulation program~\cite{gopal}.
In this version, branching fractions of heavy hadron decays 
were revised
better to reflect measured results~\cite{pdg}.
The fragmentation of b and c quarks was parametrised using
the fragmentation function of Peterson \etal~\cite{peterson},
with $\langle x_E\rangle$ 
for weakly-decaying
b and c hadrons given by the central values in Table~\ref{tab:mcpar}.
\begin{table}[htbp]
\begin{center}
\begin{tabular}{|c|c|} \hline
Quantity & Value \\ \hline
$\langle x_E\rangle_{\mathrm b}$ &$0.701\pm 0.008$ \cite{bfrag}\\
$\langle x_E\rangle_{\mathrm c}$ &$0.51\pm 0.02$ \cite{Zpaper}\\
$B (\mathrm{ b\ra\ell} )$ &$(10.5\pm 0.6 \pm 0.5)\%$ \cite{muextra}\\
$B (\mathrm{ b\ra c\ra\ell} )$ &$(7.7\pm 0.4 \pm 0.7)\%$ \cite{muextra}\\
$B (\mathrm{ b\ra \bar{c}\ra\ell} )$ 
 &$(1.3\pm 0.5)\%$ \cite{jetset,muextra}\\ \hline
\end{tabular}
 
\caption{The parameters used for the Monte Carlo simulation.
The uncertainties quoted represent the measurement errors,
as found in the quoted references.}
\label{tab:mcpar}
\end{center}
\end{table}
 
Standard Model values of the partial widths of the $\Zzero$
into $\qqbar$ were used~\cite{Stan}.
The mixture of c hadrons produced both in $\Zzero\ra\ccbar$
events and in b-hadron decays was as 
prescribed in reference~\cite{Zpaper}.
The semileptonic branching ratios of charm hadrons and associated
uncertainties were also those 
of reference~\cite{Zpaper}.
The central values in Table~\ref{tab:mcpar} were taken for the
inclusive branching ratios for $\mathrm{b\ra\ell}$,
$\mathrm{b\ra c\ra\ell}$ and $\mathrm{b \ra \bar{c} \ra \ell}$. 
The models used in describing the semileptonic decays of b and c hadrons
were those used in determining the central 
values in reference~\cite{Zpaper}.
Both $\mathrm B^{**}$ and $\mathrm D^{**}$ production were
included in the simulation, with production rates of 36\% of inclusive
B and D meson production.
Although the measured values quoted in Table~\ref{tab:mcpar}
are not in all cases the most recent, the differences
are not significant for this analysis.

Track impact parameter resolutions 
were degraded to bring the distributions
into agreement with data, as described previously~\cite{bavg}.

\section{Event selection}

The analysis was performed on data collected in the vicinity of the
$\Zzero$ peak from 1991 to 1994.  
Hadronic $\Zzero$ decays were selected using criteria
described in a previous publication~\cite{TKMH}.
For this analysis, events were rejected if the silicon microvertex
detector was not operational, resulting in a sample of
about 3.0 million events.
Tracks and electromagnetic clusters 
not associated to tracks were grouped
into jets using
a cone jet algorithm~\cite{conejet}.
The size of the cone was chosen so as to include nearly all the decay
products of a b hadron into one jet.
The jets also include some particles 
produced in the fragmentation process,
which originate directly from the 
$\mathrm{e^+e^-}$ collision point.

Electrons and muons 
with $p>2$~GeV/c were identified as described in 
reference \cite{hideto}. 
A sample enriched in semileptonic decays of b hadrons
was selected by requiring $\nn >0.7$, where $\nn$ is
the output of a neural network
based on kinematic variables \cite{hideto}.
The inputs to this network were $p$, $p_t$ and a measure
of the energy around the lepton, where
$p_t$ is the transverse momentum relative to the jet axis 
(calculated including the lepton).
The only difference compared to the previous paper \cite{hideto} is
that the requirement
$| \cos \theta | < 0.9$ was removed.
Of the final event sample, just over 3\% of the leptons have 
$| \cos \theta | > 0.9$, ranging up to $|\cos \theta| = 0.96$.

\begin{figure}[htbp]
\centering
\epsfxsize=17cm
\begin{center}
    \leavevmode
    \epsffile[30 159 532 675]{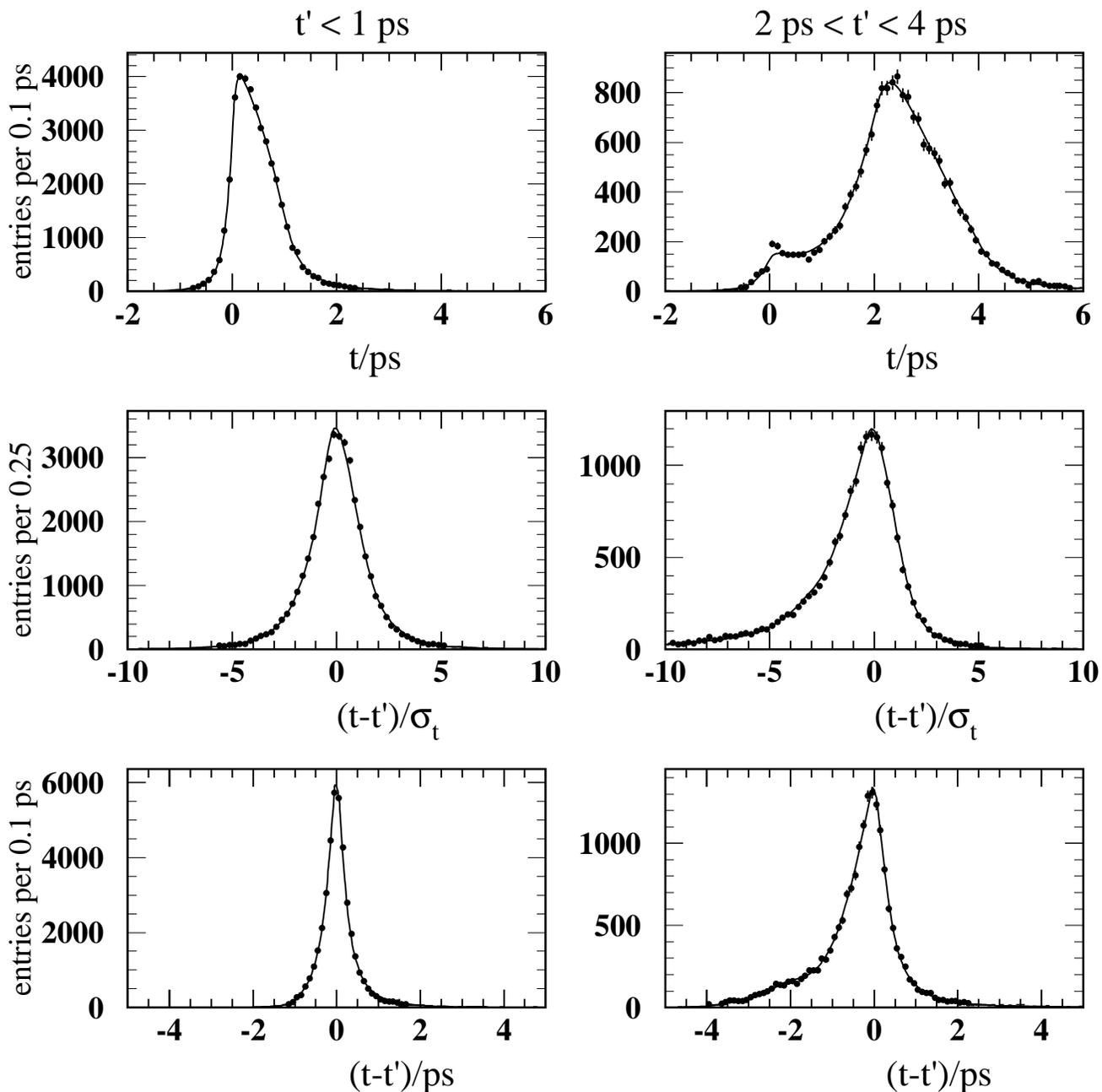}
\end{center}
\vspace{-5 mm}
\caption{ Distributions of $t$, $(t-t')/\sigma_t$ and $t-t'$
for semileptonic decays of b hadrons in two ranges of 
true decay proper times:
$t' < 1$~ps and 2~ps $< t' < 4$~ps.
The points represent the Monte Carlo simulation, and the curves
represent the parametrisation.}
\label{fig:res}
\end{figure}

\begin{figure}[htbp]
\centering
\epsfxsize=17cm
\begin{center}
    \leavevmode
    \epsffile[43 394 505 651]{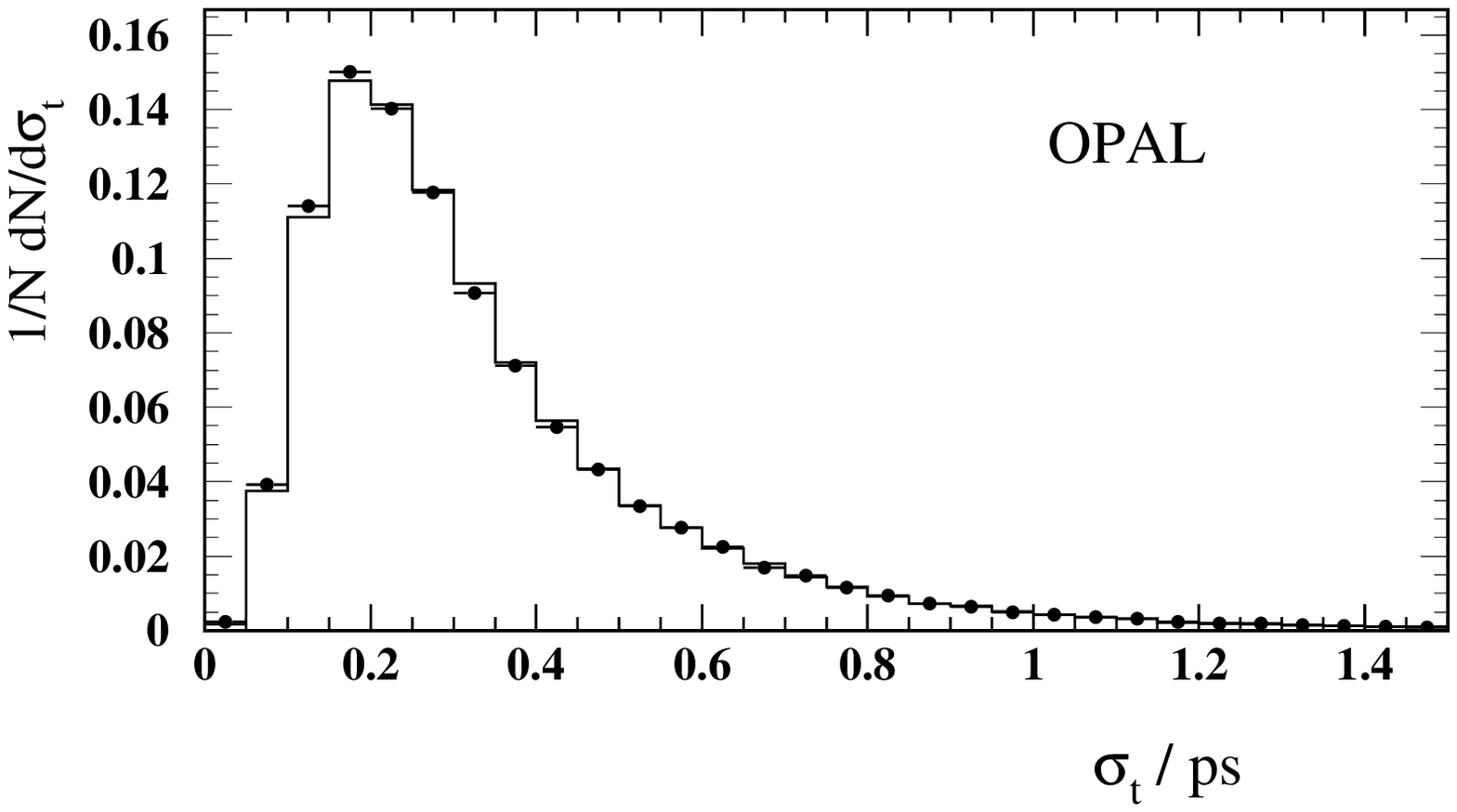}
\end{center}
\vspace{-5 mm}
\caption{ Distribution of $\sigma_t$.
The points represent the data, and the line the Monte Carlo prediction.}
\label{fig:sig}
\end{figure}

For each lepton, an estimated proper decay time  
was reconstructed for the supposed parent b-hadron 
as described previously
\cite{hideto}.
To summarise, a secondary vertex is reconstructed in the 
$x-y$ plane using
an algorithm to combine tracks with the lepton track.
The algorithm 
was able to form secondary vertices for 70\% of the lepton
candidates; the remaining 30\% were discarded.
A decay length is reconstructed by fitting to the reconstructed 
primary and secondary vertex positions, using the
jet axis as a direction constraint.
The 3-dimensional decay length is calculated by dividing the
2-dimensional decay length, $L_{xy}$, 
by the $\sin \theta$ of the jet axis.
The energy of the b hadron is estimated by 
first reconstructing the energy
of the jet that includes the lepton, using the $\Z$ mass to constrain
the event kinematics, and then subtracting an estimated contribution
from fragmentation particles.  
The fragmentation particles were separated from the 
b-hadron decay products
using momentum, angle and vertex information.
The proper time, $t$, is formed from the (3-dimensional)
decay length $L$ and boost~\footnote{
We use the notation $\hbar=c=1$.
}:
\begin{equation} t = \frac{L}{\beta \gamma} = 
 \frac{m_B}{\sqrt{E_B^2 - m_B^2}} L \; .\end{equation}
In this analysis, use is also made of the 
estimated uncertainty, $\sigma_t$,
on the proper time, calculated from the separately estimated
uncertainties on the decay length, $\sigma_L$, and the boost factor,
$\sigma_{\beta \gamma}$:
 \begin{equation} \left(\frac{\sigma_t}{t}\right)^2 =
   \left(\frac{\sigma_L}{L}\right)^2 +
   \left(\frac{\sigma_{\beta \gamma}}{\beta \gamma}\right)^2 \; , \end{equation}
where correlations between the uncertainties on $L$ and
$\beta \gamma$ are neglected.
This is unimportant because the shape of the $t$ distribution
is parametrised from Monte Carlo, as discussed in the next
section.
The use of $\sigma_t$, estimated event by event,
improves the sensitivity of the analysis
since only events with small $\sigma_t$ can discriminate between large
oscillation frequencies.
For each event, an initial estimate of $\sigma_L$ is calculated
from the error matrices of the tracks assigned to the secondary vertex.
Monte Carlo studies showed that the decay length resolution worsens
as the total momentum of tracks assigned to the vertex (excluding the
lepton) decreases.
These studies also showed that the resolution worsens with increasing
$L_{xy}$. 
Both effects result from the inclusion in the secondary
vertex of tracks that do not originate from the decay
of the b hadron or its daughters, i.e.~fragmentation products.
Neither of these effects is fully described by the initial estimate
of $\sigma_L$.
Corrections were therefore applied to $\sigma_L$ based on parametrisations
of the Monte Carlo predictions.
Monte Carlo studies of the b-hadron energy estimate showed that
the resolution depends strongly on $L_{xy}$ and the fragmentation
energy in the jet.
The uncertainty $\sigma_{\beta \gamma}$ was estimated by parametrising
the dependence on these two quantities predicted by the Monte Carlo.
%
The performance of the proper time reconstruction is shown in
Figure~\ref{fig:res} for Monte Carlo events, and the distribution
of $\sigma_t$ is shown in Figure~\ref{fig:sig}.
%
%
%

For each event, 
the b-hadron flavour at production was tagged using the
jet charge technique
described in a previous paper \cite{shlomit}.
The jet charge was defined as
\begin{equation} Q_{jet} = \sum q_i \left ( \frac{ p_i^l }{E_{\mathrm{beam}}} 
   \right )^\kappa \, , \end{equation} 
where $q_i$, $p_i^l$ are the charge and longitudinal momentum 
component (along the jet direction) of track $i$, 
$E_{\mathrm{beam}}$ is the beam energy,
and the sum is over all tracks in the jet.
The jet charge was formed for two jets: 
the jet containing the lepton
and the highest energy jet that did not contain the lepton, with
$\kappa$ chosen to be 0 and 1 respectively.
These two values of jet charge were combined into a single value,
\begin{equation} Q_{2jet} = Q_{jet}^{\kappa = 0} ( \ell ) - 
     10 \times Q_{jet}^{\kappa = 1}( \mathrm{opp} ) \; , \end{equation}
where $\ell$ denotes the lepton jet, and `opp' denotes the other jet.
The factor of 10 was chosen to give a charge separation between
produced b and $\bar{\mathrm b}$ hadrons that
is close to optimal~\cite{shlomit}.

If more than one lepton was selected in a given event, 
only the one with the highest value
of $(\frac{p}{10.})^2 + p_t^2$ was retained, where $p$ is
in GeV/c.
A total of $94\,843$ events remain after all selection criteria,
where about 75\% of the lepton candidates
are expected to originate from
semileptonic decays of b hadrons.

\section{Likelihood function}

The results obtained in this paper are extracted using
a maximum likelihood fitting procedure, where the overall
likelihood is the product of the 
likelihoods calculated for each event.
In order to construct the likelihood function,
the contributions to the selected lepton
candidates were split into 11 sources:
\begin{enumerate}
 \item lepton candidates from light quark events 
       ($\Zuu$, $\ddbar$ and $\ssbar$);
 \item leptons from semileptonic decays of c hadrons in $\Zcc$ events;
 \item lepton candidates in $\Zcc$ events not from semileptonic decays
      of c hadrons;
 \item lepton candidates in $\Zbb$ events that do not originate
       from a b hadron or its daughters;
 \item lepton candidates in $\Zbb$ events that do not come from 
       semileptonic decays of b hadrons or c hadrons, 
       but do originate from a b hadron or
       its daughters;
 \item leptons from $\bcas$ decays originating from $\Bd$
       decays;
 \item leptons from $\bcas$ decays originating from $\Bs$
       decays;
 \item leptons from $\bcas$ decays originating from $\Bp$
       or b-baryon decays;
 \item leptons from semileptonic $\Bd$ decays;
 \item leptons from semileptonic $\Bs$ decays;
 \item leptons from semileptonic $\Bp$ or b-baryon decays. 
\end{enumerate} 
Note that the notation $\bcas$ used here excludes decays of the type
$\bcb$, where the b 
refers to the quark that
decays, after any mixing has occurred.
The last three sources are
taken also to include decays of the type $\bcb$
as well as $\btau$.
Note also that decays of the type $\mathrm{b \arr J/\psi \arr \ell}$
are classed either as $\bcas$ or 
semileptonic b hadron decays 
depending on the
charge correlation of the lepton and the b quark. 
The predicted fraction of the overall sample contributed by each source
is indicated in Table~\ref{tab:frac}.
The statistical errors on these numbers are not important.
\begin{table}[htbp]
\begin{center}
\begin{tabular}{|c|ccccccccccc|} \hline
Source & 1 & 2 & 3 & 4 & 5 & 6 & 7 & 8 & 9 & 10 & 11 \\ \hline
Fraction & 0.063 & 0.089 & 0.011 & 0.005 & 0.013 & 0.025 & 0.005 & 0.025 
 & 0.299 & 0.089 & 0.377 \\ \hline
\end{tabular}
\caption{The predicted fraction of the overall selected sample due to
each source.} 
\label{tab:frac}
\end{center}
\end{table}

The joint distribution of $t$ and $\sigma_t$
is of the form
\begin{equation} 
  {\cal T}_{ij} (t, \sigma_t ) = \int {\cal S}_i (\sigma_t | t' )
   {\cal R }_i (t | \sigma_t, t' ) {\cal P }_{ij} ( t' ) 
  {\cal E}_i ( t' ) {\mathrm{d}}t' 
\end{equation}
where $t'$ is the true proper decay time for a given source,
$\cal{P}$ is the physics function, $\cal{E}$ is an efficiency function,
$\cal{ R} $ is the resolution function
for given values of $t'$ and $\sigma_t$, and $\cal{ S }$ gives the 
distribution of $\sigma_t$ for a given $t'$.
The subscript $i$ represents the source number (1 to 11). 
When CP violation is neglected, 
the subscript $j$ is 2 for leptons
from decays of B mesons that decayed as the antiparticle of the meson
produced (mixed), or 1 otherwise.
Note that the physics function describes the evolution of each
source as a function of the true proper decay time.
For sources 1, 3 and 4, $\cal{P}$ is taken to be
a delta function at $t' = 0$.
Any true lifetime content in these sources is absorbed by the
resolution function.
For source 2, $t'$ represents the proper decay time of the c hadron, 
but for all the other sources $t'$ represents the proper decay time
of the b hadron.
Sources 6,7,9 and 10 are divided into two components: 
unmixed ($j=1$) and mixed ($j=2$). 
For these components:
\begin{equation} {\cal P}_{i1} = \frac{\exp (- \frac{t'}{\tau_i})}{\tau_i} 
             \cdot \frac{1 + \cos \dmq t' }{2} \; , \;\;\;\;
   {\cal P}_{i2} = \frac{\exp (- \frac{t'}{\tau_i})}{\tau_i} 
             \cdot \frac{1 - \cos \dmq t' }{2} \; ,
\end{equation}
where 
$\dmq = \dmd \; (\dms) $, and 
$\tau_i = \tau_{\Bd} \; (\tau_{\Bs})$ for sources 6 and 9 (7 and 10).
When considering possible CP and CPT violation, sources 6 and 9
($\Bd$) were divided into four components according to the  
flavour of the B at production and decay~\cite{alrick} :
\begin{eqnarray}
 {\Bd \arr \Bd} \;\; : \;\; 
 {\cal P}_{i1} & = &\frac{\exp (- \frac{t'}{\tau_i})}{\tau_i} \cdot 
  \frac{1 + \cos \dmd t' - 4 \idel \sin \dmd t'}{2 k_{\mathrm B}}
  \nonumber \\
 {\Bd \arr \bar{\Bd}} \;\; : \;\; 
 {\cal P}_{i2} & = & \frac{\exp (- \frac{t'}{\tau_i})}{\tau_i} \cdot
  \frac{1 - \cos \dmd t' }{2} \cdot \frac{1 - 4\reps}{k_{\mathrm B}}
  \nonumber \\
{\bar{\Bd} \arr \bar{\Bd}} \;\; : \;\; 
 {\cal P}_{i3} & = & \frac{\exp (- \frac{t'}{\tau_i})}{\tau_i} \cdot 
  \frac{1 + \cos \dmd t' + 4 \idel \sin \dmd t'}{2 k_{\mathrm
      \bar{B}}} \nonumber \\
 {\bar{\Bd} \arr \Bd} \;\; : \;\; 
 {\cal P}_{i4} & = & \frac{\exp (- \frac{t'}{\tau_i})}{\tau_i} \cdot
  \frac{1 - \cos \dmd t' }{2} \cdot \frac{1 + 4\reps}{k_{\mathrm \bar{B}}}
\end{eqnarray}
where $k_{\mathrm B}$ is a constant which ensures the normalisation 
$\int ({\cal P}_{i1} + {\cal P}_{i2}) {\mathrm d}t' = 1$, 
and similarly for $k_{\mathrm \bar{B}}$ and 
${\cal P}_{i3} + {\cal P}_{i4}$.
Note that CP and CPT violation in the $\Bs$ system is neglected
in this analysis.
Sources 8 and 11 were described by a double exponential, with the 
lifetime parameters
corresponding to the $\Bp$ and b-baryon lifetimes.
Source 2 was also parametrised by a double exponential to describe the
different individual lifetimes of the c hadrons.
This approximation gave a good description of 
this background in the Monte Carlo.

The function ${\cal E}$ describes the efficiency for reconstructing
a proper decay time as a function of $t'$. 
A single function was common for sources 9,10,11, a second function
was common for sources 6,7,8, and individual functions were used
for sources 2 and 5.
For sources 9,10,11,
the efficiency is about 5\% smaller at $t'=0$~ps than at $t'=1.5$~ps,
above which point it is flat.

The functions $\cal S$ and $\cal R$ were parametrised using Monte Carlo
events. 
For this purpose, one set of functions was common to sources 5,9,10,11,
and another set was common to sources 6,7,8.
The effects of leptons from secondary decays
($\bcas$, $\bcb$, $\btau$) are taken into account in the
resolution functions, rather than the physics functions.
Distributions of $t$, $t-t'$ and $(t-t')/\sigma_t$
for semileptonic decays of b hadrons 
are shown in 
Figure~\ref{fig:res} 
for two ranges of $t'$.
These are compared to the resolution function $\cal R$ integrated over
the relevant proper time interval,
and also over $\sigma_t$. 
The distribution of $\sigma_t$ is shown for data and Monte Carlo in
Figure~\ref{fig:sig},
equivalent to $\cal S$ integrated over $t'$ and averaged over 
all sources.
The Monte Carlo describes the data well, and it is clear
that some events are measured much better than others.

The normalisation of the joint distribution function
$\cal T$ for each source is determined by
\begin{equation} 
 \int \sum_{j} {\cal P }_{ij} ( t' ) {\cal E}_i ( t' ) {\mathrm{d}}t' = 1 \; , \;\;
   \int {\cal R }_i (t | \sigma_t, t' ) {\mathrm{d}}t = 1 \; , \;\; 
   \int {\cal S}_i (\sigma_t | t' ) {\mathrm{d}}\sigma_t = 1 \; . \end{equation}
For sources involving
mixing, the normalisation condition for $\cal P$ applies
to the sum over mixed and unmixed functions.

\begin{figure}[htbp]
\centering
\epsfxsize=17cm
\begin{center}
    \leavevmode
    \epsffile[42 395 503 652]{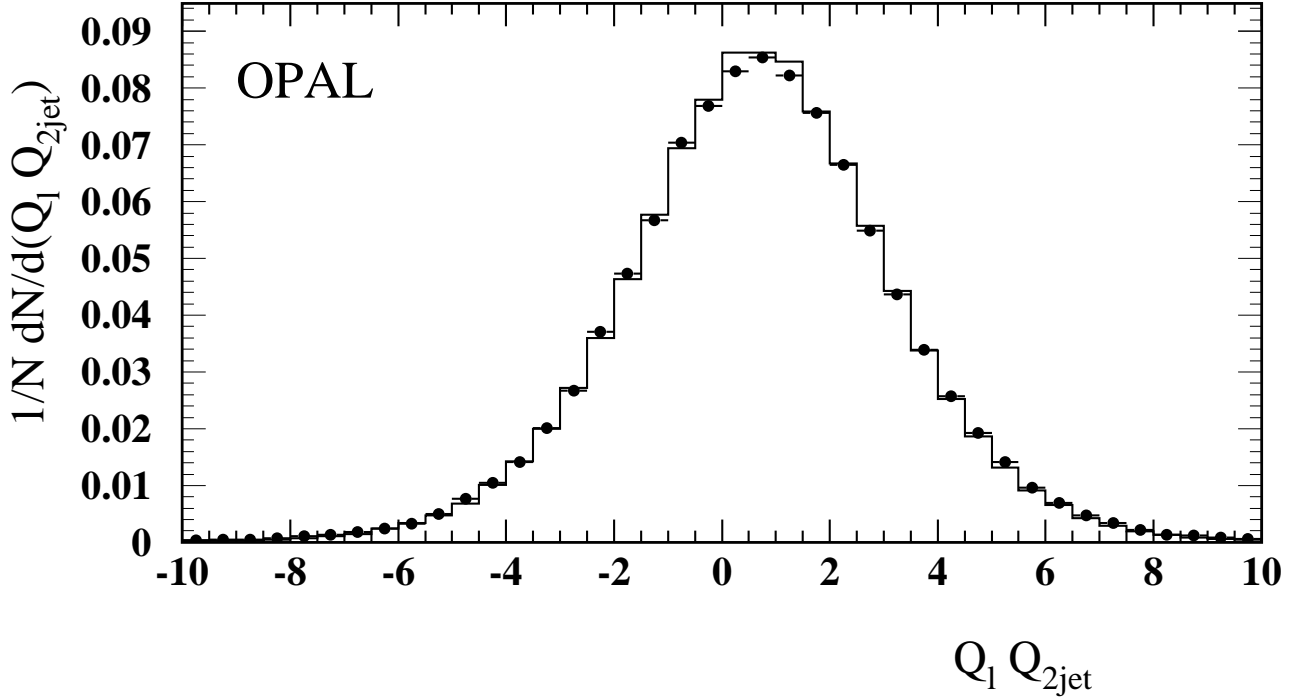}
\end{center}
\vspace{-5 mm}
\caption{ The distribution of $Q_{\ell} Q_{2jet}$ for data is 
shown by the points with error bars.
The curve indicates the Monte Carlo prediction, assuming
$\dmd = 0.45$~ps$^{-1}$
and  $\dms = 15$~ps$^{-1}$. }
\label{fig:q}
\end{figure}

\begin{figure}[htbp]
\centering
\epsfxsize=17cm
\begin{center}
    \leavevmode
    \epsffile[42 395 503 652]{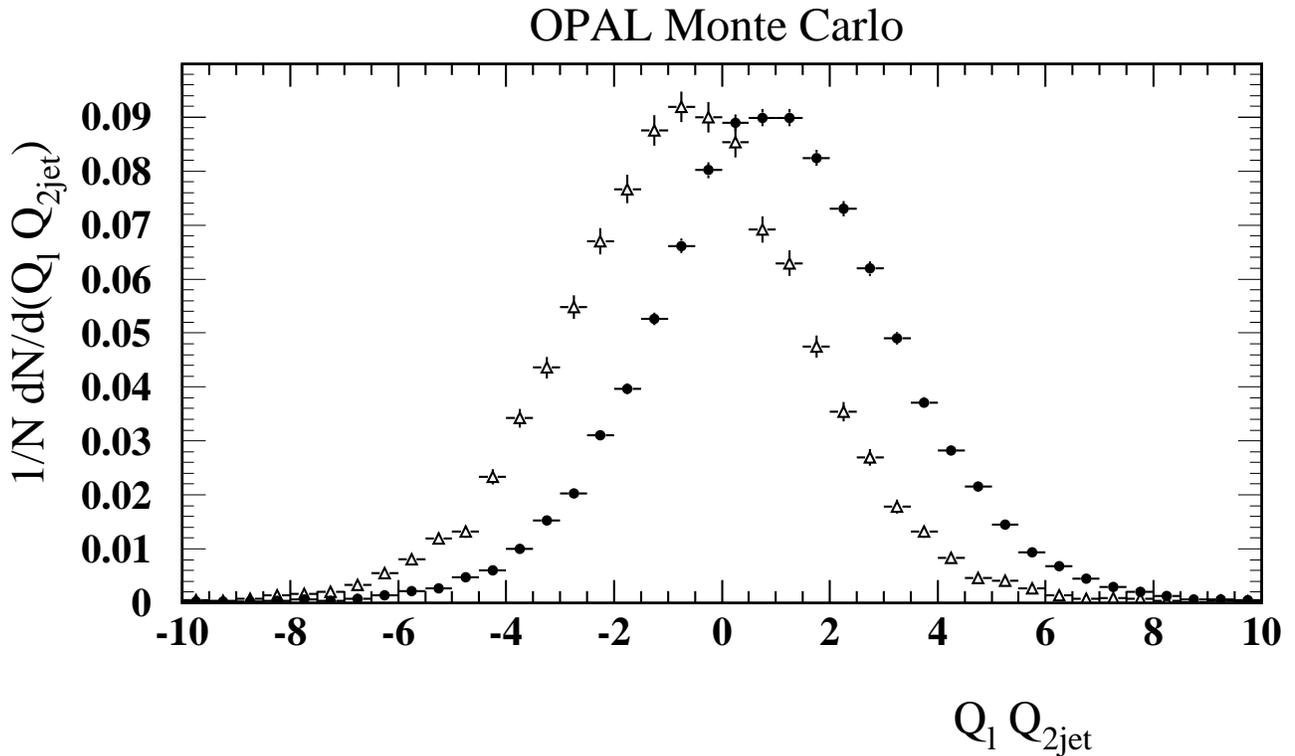}
\end{center}
\vspace{-5 mm}
\caption{ The
distribution of $Q_{\ell} Q_{2jet}$ for semileptonic decays
of neutral B mesons in the Monte Carlo. 
The solid points represent unmixed B mesons while the triangles
show the distribution for mixed B mesons.}
\label{fig:mq}
\end{figure}

To describe the charge distributions for each source, we first define
the fraction of each source, $\xi_{ij}$, for which 
the lepton has the same charge sign as the quark,
at its production, 
that produced the lepton jet.
For example $\xi$ is 1 for semileptonic decays of b hadrons, but 0 for
$\bcas$ decays when no mixing has occurred in either case.
The full set of values for $\xi$ is given in Table~\ref{tab:xi},
where the values for sources 1,3,4 and 5 are taken from the
Monte Carlo simulation.
Variation of these values was found to lead to negligible 
changes in the results of this paper.
The sign of the jet charge, $Q_{2jet}$, is an estimate of the charge
sign of the quark, before any mixing, that produced the lepton jet.
The quantity $\eta_{ij}( |Q_{2jet}| )$ is defined as 
the probability that these 
charge signs disagree, where $i$ represents the source number and
$j=1$ or 2
(also 3 or 4 when considering CP(T) violation
as in equation 10)
to represent no mixing or mixing, respectively, in the lepton jet.
Monte Carlo events were used to parametrise $\eta$ as a 
function of $|Q_{2jet}|$.
For sources from $\Zbb$ events, i.e. sources 4 to 11,
the effect of mixing in the jet that does not contain the lepton was
taken into account by modifying $\eta$ as a function of $\chi$, the 
time-averaged mixing parameter, averaged over all b hadron species
weighted by their production fractions in $\Z$ decays:
\begin{equation}
 \eta_{ij}^{ \chi } = \eta_{ij}^{ \chi = 0 } \cdot ( 1 - X ) +
                  (1 - \eta_{ij}^{ \chi = 0 } ) \cdot X \; , 
 \;\; {\mathrm{where}} \;\; X = \chi  \cdot w( |Q_{2jet}| ) \; , 
\end{equation}
and the weighting function $w( |Q_{2jet}| )$ was parametrised 
from Monte Carlo data.
Note that $\chi$ is computed in the fit from $\dmd$, $\dms$, the
lifetimes and the fractions of $\Bd$ and $\Bs$ mesons produced.
To describe the charge distribution for each source, we define
\begin{equation}
{\cal Q}_{ij} = \left\{ \begin{array}{ll}
   \xi_{ij} \cdot (1 - \eta_{ij} ) + ( 1-\xi_{ij} ) \cdot \eta_{ij} 
             & \mbox{if $Q_{\ell} Q_{2jet} > 0$ } \\
   (1-\xi_{ij} ) \cdot (1-\eta_{ij} ) + \xi_{ij} \eta_{ij} 
             & \mbox{otherwise} \\ \end{array} \right. \; ,\end{equation}
where $Q_{\ell}$ is the charge of the lepton candidate.
The function ${\cal Q}_{ij}$ represents the probabilities
for finding $Q_{\ell} Q_{2jet}$ positive or negative
for a given $|Q_{2jet}|$.
A comparison of the distribution of $Q_{\ell} Q_{2jet}$
between data and Monte Carlo is shown in Figure~\ref{fig:q},
where the Monte Carlo distribution assumed $\dmd = 0.45$~ps$^{-1}$
and  $\dms = 15$~ps$^{-1}$. 
A reasonable agreement is observed.
The distributions of $Q_{\ell} Q_{2jet}$ for semileptonic
decays of neutral B mesons that have and have not mixed are
compared in Figure~\ref{fig:mq}.
For neutral unmixed B mesons, $\eta$ is about 0.3 for
$| Q_{2jet} | = 2$.
When considering CP(T) violation, two out of the four physics
functions defined in equation 10
are used for a particular source, according to the lepton
charge.
For example, for a negatively charged lepton, the functions
corresponding to $j=1$ and $j=4$ would be used for source 8.
\begin{table}[htbp]
\begin{center}
\begin{tabular}{|cc|ccccccccccc|} \hline
Source & & 1 & 2 & 3 & 4 & 5 & 6 & 7 & 8 & 9 & 10 & 11 \\ \hline
$\xi$ & unmixed & 0.69 & 1 & 0.62 & 0.67 & 0.67 & 0 & 0 & 0 & 1 & 1 & 1 \\ 
      & mixed &   &   &   &   &   & 1 & 1 &   & 0 & 0 &   \\ \hline
\end{tabular}
\caption{Values of $\xi$ for each source. The second row represents
the case of leptons from B mesons that have mixed.}
\label{tab:xi}
\end{center}
\end{table}
%
%
%

\begin{figure}[htbp]
\centering
\epsfxsize=17cm
\begin{center}
    \leavevmode
    \epsffile[41 385 502 652]{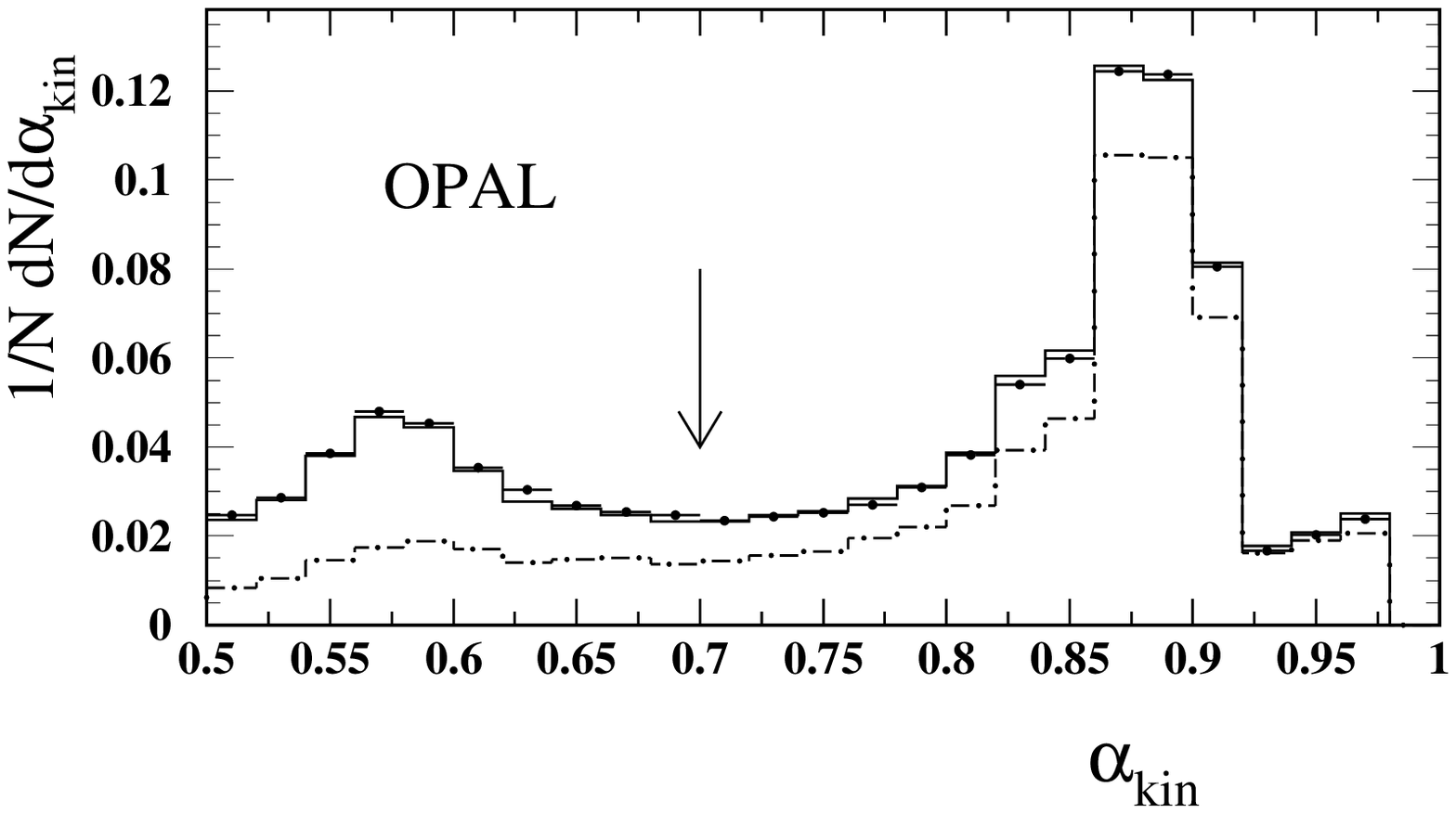}
\end{center}
\vspace{-5 mm}
\caption{ The distribution of $\nn$ for data is 
shown by the points.
The solid histogram shows the Monte Carlo prediction, and
the dashed-dotted histogram shows the contribution from
semileptonic decays of b hadrons.
The arrow indicates the selection cut; events to the left
of the arrow were rejected.}
\label{fig:nn}
\end{figure}

The final ingredient to the likelihood is the fraction
of each source, which is estimated event by event based on the value
of the neural net output, $\nn$, with a small 
correction for the absolute value of the jet charge,
$|Q_{2jet}|$, giving
\begin{equation}
 {\cal N }_i(\nn,|Q_{2jet}|) = 
\frac{ {\cal A}_i(\nn) {\cal B}_i (|Q_{2jet}|) } 
{ \sum_i {\cal A}_i(\nn) {\cal B}_i (|Q_{2jet}|) } \; .
\end{equation}
The function
${\cal A}_i$ is the fraction of events due to
source $i$ for a given value of $\nn$, 
and ${\cal B}_i$ describes the distribution
$|Q_{2jet}|$ for source $i$. 
Note that different sources give rise to slightly different
distributions of $|Q_{2jet}|$, 
in particular light quark events give rise to larger
values of $|Q_{2jet}|$ than do $\bbbar$ events on average.
The correction, using the function ${\cal B}_i$, takes this into account.
Sources 9, 10 and 11 give rise to very similar distributions of $|Q_{2jet}|$,
and are described by a common function ${\cal B}$.
The function ${\cal A}_i (\nn)$ 
is parametrised from Monte Carlo events, taking into account
the recipe for modelling the momentum spectra of leptons in the
rest frame for the semileptonic decays of b and c hadrons 
and for $\bcas$
decays described in a previous paper \cite{Zpaper}.
The predicted distribution of $\nn$ is compared to the data in
Figure~\ref{fig:nn}, where the contribution of events from
semileptonic decays of b hadrons is also indicated.
Note that the semileptonic branching ratios of the
individual b hadrons were assumed to be proportional to
the lifetimes.

The overall likelihood of the event sample is taken as
\begin{equation}
 {\cal L}_{\mathrm{data}} = 
   \prod_{m=1}^{n} \sum_{i, j} {\cal T}_{ij} (t^m, \sigma_t^m )
             {\cal Q}_{ij} ( Q_{\ell}^m Q_{2jet}^m )
             {\cal N }_i(\nn^m,|Q_{2jet}^m|) \; ,
\end{equation}
where the product is over all the events in the sample.
 
\section{Results}

In order to obtain information about $\dmd$ and $\dms$,
the likelihood, ${\cal L}$,
was maximised with respect to $\dmd$ for many
values of $\dms$.
However, in order to include systematic uncertainties,
other parameters in the likelihood calculation were
also allowed to vary, but under Gaussian constraints, i.e.
\begin{equation}
{\cal L} = {\cal L}_{\mathrm{data}} \times {\cal L}_{\mathrm{constr}} \;
\end{equation}
where ${\cal L}_{\mathrm{constr}}$ is the product of the Gaussian constraints.
Thus a multiparameter fit was performed for each $\dms$ point.
This procedure was also employed
in reference~\cite{hideto}.
The parameters that were allowed to vary and the constraints
on these parameters
are listed in Table~\ref{tab:sys}.
\begin{table}[htbp]
\begin{center}
\begin{tabular}{|c|c|} \hline
Parameter & Constraint \\ \hline
$f_{\mathrm s}$ & see text \\ 
$f_{\mathrm{baryon}}$ & $(13.2 \pm 4.1)$\% \\
$f_{\bcas}$ & $\pm 15$\% $\times$ nominal \\
$f_{\mathrm c}$ & $\pm 20$\% $\times$ nominal \\
$f_{\mathrm{uds}}$ & $\pm 20$\% $\times$ nominal \\
 & \\
$\langle \tau_{\mathrm b} \rangle $ & $1.549 \pm 0.020$ ps \\
$\tau^+ / \taud $ & $ 1.03 \pm 0.06 $ \\
$\tau_{\mathrm s} / \taud $ & $ 1.03 \pm 0.075 $ \\
$\tau_{\Lb} / \taud $ & $ 0.73 \pm 0.06 $ \\
 & \\
$\delta Q_{\mathrm b}$ & $0 \pm 0.06$ \\
$\delta Q_{\mathrm{mix}}$ & $0 \pm 0.06$ \\
$\delta Q_{\Bp}$ & $0 \pm 0.06$ \\
$\delta Q_{\mathrm{udsc}}$ & $0 \pm 0.06$ \\
$f_{\mathrm D^{**}}$ & $(35 \pm 10)$\% \\ \hline
\end{tabular}
\caption{Constrained parameters in the fit} 
\label{tab:sys}
\end{center}
\end{table}
In this table, $f_{\mathrm s}$ is defined as the production fraction
$f(\mathrm b \arr \Bs)$.
The value is constrained both by direct
measurements, giving a rate of $(11.1 \pm 2.6)$\% \cite{pdg} relative to
all weakly decaying b hadrons, and by the measured average mixing rate
of b hadrons, $\chi = 0.126 \pm 0.008$ \cite{pdg} together with knowledge
of the equivalent parameters, $\chi_{\mathrm d(s)}$, for $\Bd$($\Bs$)
mesons and the fraction of b baryons,
where
\begin{equation}
 \chi_{\mathrm d}=\frac{1}{2} \left ( \frac{x_{\mathrm d}^2}{1+x_{\mathrm d}^2} 
 \right ) \; ,
  \;\;\;\; x_{\mathrm d} = \dmd  \taud \;,
\end{equation}
with $\taud$ representing the $\Bd$ lifetime,
and similarly for $\chi_{\mathrm s}$.
This is equivalent to the constraint 
$f_{\mathrm s} = (11.2^{+1.8}_{-1.9})$\% \cite{pdg}
except that the values of $\chi_{\mathrm d}$ and $\chi_{\mathrm s}$
are calculated from the values of $\dmd$ and $\dms$ in the fit,
together
with the appropriate lifetimes,
and  $f_{\mathrm{baryon}}$, defined as 
$f(\mathrm b \arr $ b baryon), is also taken from the value in the fit.
The quoted constraint on $f_{\mathrm{baryon}}$ is taken
from reference~\cite{pdg}.
When varying $f_{\mathrm s}$ and $f_{\mathrm{baryon}}$, the following
rules are applied:
\begin{equation}
f_{\mathrm d} + f_{\mathrm u} + f_{\mathrm s} + f_{\mathrm{baryon}} =
1 \;\; {\mathrm{and}} \;\; f_{\mathrm d} = f_{\mathrm u} \; ,
\end{equation}
where $f_{\mathrm d}$ and $f_{\mathrm u}$ are the equivalent
production fractions for $\Bd$ and $\Bp$ respectively.
The actual fraction of selected leptons coming from 
$\Bs$ mesons (b baryons) is also affected by
the semileptonic branching ratio, which is assumed to be proportional
to the ratio of lifetimes
$\tau_{\mathrm s} / \langle \tau_{\mathrm b} \rangle $
($\tau_{\Lb} / \langle \tau_{\mathrm b} \rangle $),
where $\tau_{\mathrm s}$ is the $\Bs$ lifetime,
$\tau_{\Lb}$ the average lifetime for b baryons
and $\langle \tau_{\mathrm b} \rangle $ is the average lifetime
for b hadrons.
Similarly, the fraction of selected leptons coming from $\Bp$ relative
to $\Bd$ decays depends on the ratio $\tau^+ /\taud$, where $\tau^+$
is the $\Bp$ lifetime.
The parameter $f_{\bcas}$ is a scaling factor affecting the source fractions
for $\bcas$ processes, sources 6 to 8.
The uncertainty quoted includes uncertainties 
due to branching fractions,
decay modelling and detector simulation.
The parameter $f_{\mathrm c}$ is a scale factor multiplying the source
fractions for sources 2 and 3, and $f_{\mathrm{uds}}$ is
the corresponding parameter for source 1.
The uncertainty on $f_{\mathrm c}$ covers uncertainties 
due to the charm semileptonic branching ratio, the modelling
of the $\nn$ distribution and lepton background.
The uncertainty on $f_{\mathrm{uds}}$ is due to uncertainties
in the lepton background rates.
The lifetime constraints indicated in the table were taken from
the Particle Data Group review~\cite{pdg}. 
Note that the existence of two different lifetimes for the $\Bs$ mass
eigenstates is expected to have a negligible effect on this analysis
and has been neglected.

The parameter $\delta Q_{\mathrm b}$ represents an offset to the 
Monte Carlo distributions
of $Q_{2jet}$ for $\Zbb$ events before calculation of $\eta_{ij}$.
Such an offset moves the distributions for produced b and
$\mathrm{\bar{b}}$ hadrons closer together or further apart.
Thus it causes a change in $\eta$ as a function of $Q_{2jet}$, and is
equivalent to an uncertainty in $\eta$ of about $\pm 0.02$ when 
$\eta = 0.25$.
Monte Carlo predictions for 
$\eta$ were tested to this
level in a previous paper~\cite{shlomit}.
The parameter $\delta Q_{\mathrm{mix}}$ has the same definition
as $\delta Q_{\mathrm b}$, but applies only to events where the lepton
originates from a mixed B meson.
The parameter $\delta Q_{\Bp}$ is also defined in the same way,
but applies only to events where the lepton originates from $\Bp$
decays.
Similarly, $\delta Q_{\mathrm{udsc}}$ applies only to 
$\uubar$, $\ddbar$, $\ssbar$
or $\ccbar$ events.
The parameter $f_{\mathrm D^{**}}$ simulates the 
uncertainty 
on the rate of $\bcas$ decays from $\B0$ and $\Bp$
mesons relative to all b hadrons
due to the effect of 
varying the rate of decays through a 
$\mathrm D^{**}$. 
A variation in the rate of $\bcas$ decays from $\Bs$ mesons
relative to all b hadrons was also considered, but found to have a
negligible effect.

The uncertainty due to the resolution functions, $\cal{R}$ and $\cal{S}$,
was not taken into account using this technique.
Instead, these functions were 
reparametrised without applying the smearing of the track impact parameters
in the Monte Carlo, 
and the 
likelihood analysis repeated using the new functions.
The effect of fragmentation uncertainties on the resolution functions
was assessed by using Monte Carlo events generated with 
$\langle x_E\rangle_{\mathrm b}$ 
shifted by 0.02 relative to the central value.
Such a change represents a shift of over $2\sigma$ 
with respect to the measured value~\cite{bfrag}, but is inflated
to include the effect of shape uncertainties.
Uncertainty due to charm fragmentation is expected
to have a negligible effect, and
was neglected.

\subsection{Determination of \mbox{\boldmath $\dmd$} }

\begin{figure}[htbp]
\centering
\epsfxsize=17cm
\begin{center}
    \leavevmode
    \epsffile[21 157 531 649]{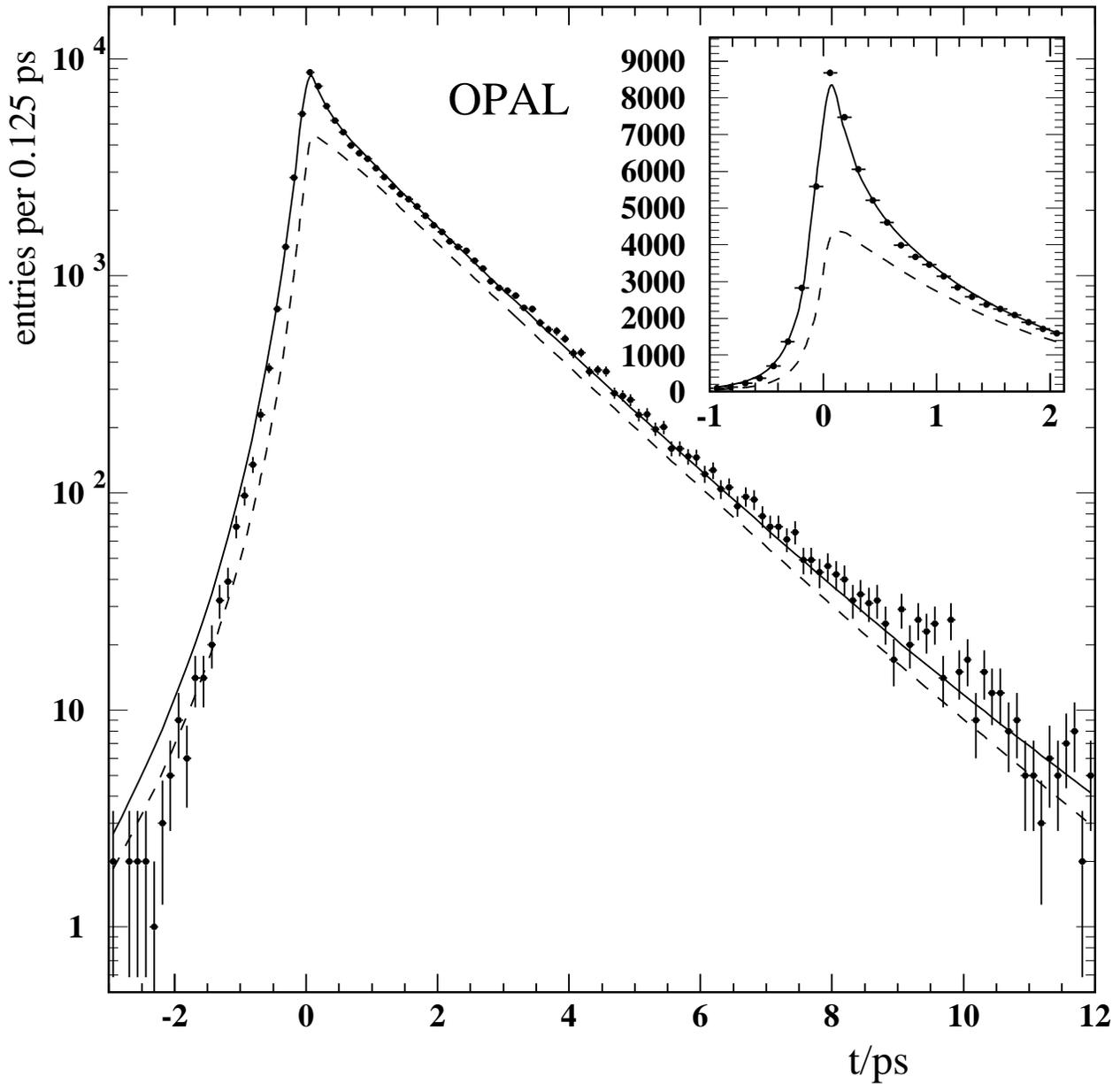}
\end{center}
\vspace{-5 mm}
\caption{ 
The distribution of $t$ for the selected events in the data.
The solid line represents the result of the fit, while the dashed line
indicates the component from semileptonic decays of b hadrons, sources
9, 10 and 11. 
The insert shows the same distribution on a linear scale.
}
\label{fig:t}
\end{figure}

With $\dms$ fixed to 15~ps$^{-1}$ the multiparameter fit was performed,
with the result
\[ \dmd = 0.444 \pm 0.034~\mathrm{ps^{-1}} \; . \]
This error includes systematic components, due to the constraints
of Table~\ref{tab:sys}, as well as a statistical component.
The fitted value of $f_{\mathrm s}$ was $(10.6 \pm 1.9)$\%, and
the calculated value of $\chi$ was 0.115. 
The distribution of $t$ from the fit is superimposed on the data
in Figure~\ref{fig:t}.
%
%
%
\begin{figure}[htbp]
\centering
\epsfxsize=17cm
\begin{center}
    \leavevmode
    \epsffile[21 399 531 652]{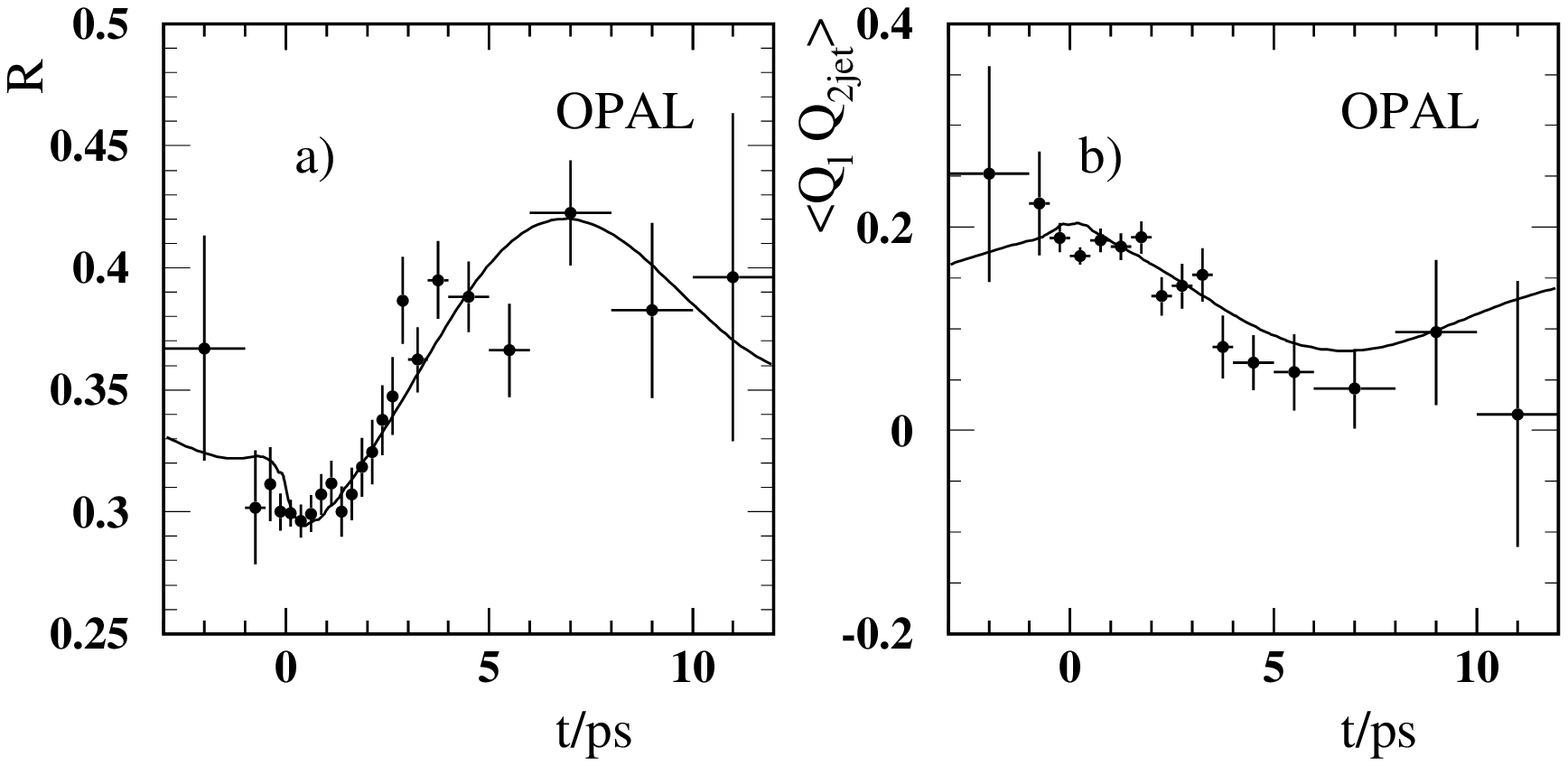}
\end{center}
\vspace{-5 mm}
\caption{ 
a) The ratio $R$ versus $t$ 
for data with $| Q_{2jet} | > 2$. 
b) 
$\langle Q_{\ell} Q_{2jet} \rangle$
versus $t$ for data with
$| Q_{2jet} | < 2$.
The line shows the result of the fit when $\dms = 15$~ps$^{-1}$ 
in both a) and b).
}
\label{fig:tr2}
\end{figure}
Shown in Figure~\ref{fig:tr2} is the ratio 
$R$ versus $t$ for $| Q_{2jet} | > 2$, and 
$\langle Q_{\ell} Q_{2jet} \rangle$ versus $t$ for
$| Q_{2jet} | < 2$.
Here $R$ is defined as the fraction of events that have 
$Q_{\ell} Q_{2jet} < 0$.
The ratio $R$ is not an appropriate variable to plot
for $| Q_{2jet} | < 2$, because $\eta$ changes rapidly with 
$| Q_{2jet} |$ in this region and, in particular, many of the
events have $| Q_{2jet} |$ close to 0, where there is no sensitivity.
The mean value of $Q_{\ell} Q_{2jet}$ is more sensitive
than $R$ in this region.
Conversely, since $\eta$ changes only slowly with $Q_{2jet}$
when $| Q_{2jet} | > 2$, $R$ is the more appropriate variable in
this region.

The systematic error on $\dmd$ due to tracking resolution was assessed,
as described in the previous section, to be 
$\pm 0.004$~ps$^{-1}$.
The contribution from the b fragmentation uncertainty
was found to be $\pm 0.003$~ps$^{-1}$.
The effect of the uncertainty in the fraction of B mesons
decaying to $\mathrm D^{**}$ on the fitted resolution functions
was assessed by repeating the analysis using resolution functions
determined from a Monte Carlo that did not include $\mathrm D^{**}$.
The difference between the resulting $\dmd$ value and the central
value quoted above was divided by 3 to correspond to an uncertainty
of $\pm 10\%$ (a relative error of 30\%)
in the production of $\mathrm D^{**}$ per B meson.
This procedure contributed an uncertainty of $\pm 0.001$~ps$^{-1}$ on
$\dmd$. 
The systematic error due to $\dms$ was assessed by taking the
largest excursions on the fitted value of $\dmd$ when the fit
was repeated for different values of $\dms$ in the range 
$\dms \geq 6.6$~ps$^{-1}$.
We consider the range $\dms < 6.6$~ps$^{-1}$ to be 
excluded by previous
measurements~\cite{hideto,aleph,Wu,dsl}.
The error on $\dmd$ from this source 
was $^{+0.009}_{-0.0002}$~ps$^{-1}$.
The errors on $\dmd$ resulting from different ranges of $\dms$
are given in Table~\ref{tab:dmddms}.
The uncertainty due to the weighting function $w( |Q_{2jet}| )$
was found to be negligible.
\begin{table}[htbp]
\begin{center}
\begin{tabular}{|c|c|} \hline
Lower limit on $\dms$ (ps$^{-1}$) & 
Systematic uncertainty on $\dmd$(ps$^{-1}$) \\ \hline
3  &  $+0.014 \;$  $\; -0.0002$ \\
6.6 & $+0.009 \;$  $\; -0.0002$ \\
8  &  $+0.004 \;$  $\; -0.0002$ \\
10 &  $+0.001 \;$  $\; -0.0002$ \\ \hline
\end{tabular}
\caption{The systematic error on $\dmd$ resulting from different
ranges of $\dms$ values.} 
\label{tab:dmddms}
\end{center}
\end{table}

The final result for $\dmd$ is 
$\dmd =  0.444 \pm 0.034~^{+0.011}_{-0.005}~\mathrm{ps}^{-1}$,
where, as above, the first error has a systematic component.
To aid comparison with previous measurements, the systematic
components to the fit error 
(due to the constraints of Table~\ref{tab:sys})
were calculated by determining the
effect of changing each constraint
by the error given in Table~\ref{tab:sys}, 
one at a time.
The results are shown in
Table~\ref{tab:sys2}.
\begin{table}[htbp]
\begin{center}
\begin{tabular}{|c|c|} \hline
Origin & Systematic error on $\dmd$ in ps$^{-1}$ \\ \hline
$f_{\mathrm s}$ & $^{-0.011}_{+0.010}$ \\
$f_{\mathrm{baryon}}$ & $^{+0.005}_{-0.004}$ \\
$f_{\bcas}$ & $\mp 0.002$ \\
$f_{\mathrm c}$ & $\pm 0.000$ \\
$f_{\mathrm{uds}}$ & $\pm 0.000$ \\
 & \\
$\langle \tau_{\mathrm b} \rangle $ & $\pm 0.000$ \\
$\tau^+ / \taud $ & $\pm 0.004 $ \\
$\tau_{\mathrm s} / \taud $ & $\mp 0.002 $ \\
$\tau_{\Lb} / \taud $ & $^{+0.007}_{-0.006} $ \\
 & \\
$\delta Q_{\mathrm b}$ & $\mp 0.001$ \\
$\delta Q_{\mathrm{mix}}$ & $\mp 0.007$ \\
$\delta Q_{\Bp}$ & $\pm 0.006$ \\
$\delta Q_{\mathrm{udsc}}$ & $\mp 0.003$ \\
$f_{\mathrm D^{**}}$ & $\pm 0.000$ \\ 
 & \\
Tracking resolution & $\pm 0.004$ \\
b fragmentation & $\pm 0.003$ \\
Resolution effect of $\mathrm D^{**}$ & $\mp 0.001$ \\
$\dms$ variation & $^{+0.009}_{-0.000}$ \\ \hline
Total & $^{+0.020}_{-0.017}$ \\ \hline
\end{tabular}
\caption{Summary of systematic errors on $\dmd$} 
\label{tab:sys2}
\end{center}
\end{table}
We can now rewrite the result for $\dmd$ as
\[
\dmd = 0.444 \pm 0.029~^{+0.020}_{-0.017}~\mathrm{ps}^{-1} \; , 
\]
where the first error is essentially statistical 
and the second is systematic.
Note that all constrained parameters were consistent with the 
constraints to within one standard deviation, except
$f_{\bcas}$, $f_{\mathrm{uds}}$ and $\delta Q_{\mathrm{udsc}}$
where the fitted values were
larger than
the constraint values, but were consistent at the level of
two standard deviations. 

\subsection{Study of \mbox{\boldmath $\dms$} }

\begin{figure}[htbp]
\centering
\epsfxsize=17cm
\begin{center}
    \leavevmode
    \epsffile[20 150 525 674]{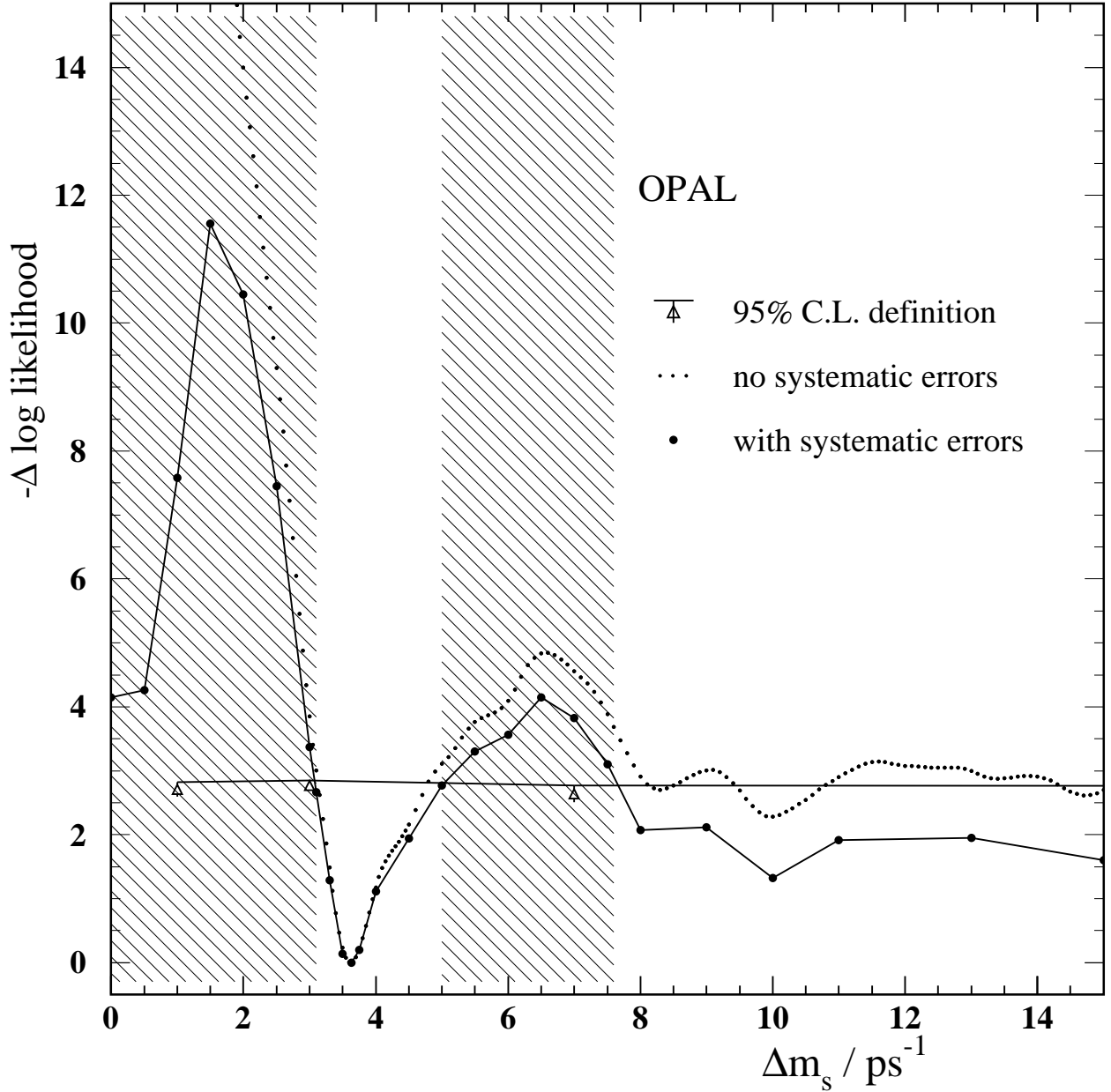}
\end{center}
\vspace{-5 mm}
\caption{ 
The dotted line shows the distribution of $- \Delta \log \cal L$ with respect
to the maximum value of $\log \cal L$ as a function of $\dms$, while
keeping all other parameters fixed to values optimised at 
$\dms = 3.63$~ps$^{-1}$.
The solid points include the effects of systematic uncertainties.
The open triangles 
represent the position of the 95\% confidence level, and the
horizontal line joins together these points, increased by one
standard deviation. 
The hatched region indicates values of $\dms$ excluded at 95\%
confidence level by this analysis.
}
\label{fig:lik}
\end{figure}

In order to constrain $\dms$,
we consider two different techniques.
For the first technique,
we plot in Figure~\ref{fig:lik} the value of $- \Delta \log \cal L$
as a function of $\dms$, 
where $\Delta \log \cal L$ indicates
the difference in $\log \cal L$ relative to the maximum value
of $\log \cal L$.
%
%
%
The solid points include the effects of the systematic uncertainties.
They were obtained by allowing all parameters,
including $\dmd$, to vary at each value of $\dms$.
To account for systematic errors due to the description of
the resolution,
four different sets of resolution functions were used.
In addition to the default functions, we used the resolution functions
obtained without the impact parameter smearing
in the Monte Carlo, and those obtained using the Monte Carlo
with different fragmentation parameters.
A fourth choice of resolution function was obtained by 
using only semileptonic $\Bs$ decays to parametrise sources
5,9,10 and 11 (using the nominal tracking resolution).
For each of the four sets of resolution functions
the full minimisation of the other parameters was performed
at each $\dms$ point, giving four distributions of $\Delta \log \cal L$,
with respect to the point of maximum $\cal L$ relevant to each set,
versus $\dms$.
The solid points were obtained simply by taking the minimum value
of $\Delta \log \cal L$ from the four choices at each $\dms$ point.
As a result of this procedure,
the points for $\dms < 1$~ps$^{-1}$ and for $\dms \geq 3.6$~ps$^{-1}$
are taken
from the resolution functions without the impact parameter smearing,
while those for 1~ps$^{-1} \leq \dms < 3.6$~ps$^{-1}$ come from the
resolution functions obtained with the modified fragmentation
parameters.
The value of $\Delta \log \cal L$ at $\dms=\infty$ was found to be 2.6
(including systematic errors).
%
%
A limit on $\dms$ is extracted by finding
the intersection of these points 
with a limit curve calculated using a Monte Carlo
technique. 
This technique consisted of performing fast simulations
of the data sample, assuming various values of $\dms$.
The simulated samples were then analysed in the same way 
as the data (including the multiparameter fit), and a distribution of
$\Delta \log \cal L$ obtained at each $\dms$ point.
The limit curve is defined by the value of $\Delta \log \cal L$
above which lie only 5\% of the simulated samples.
A total of 6000 samples were simulated, distributed over four values
of $\dms$.
Three of the limit points and the limit curve,
obtained by linear interpolation, are indicated in
Figure~\ref{fig:lik}. 
The fourth limit point is at $\dms=50$~ps$^{-1}$.

We extract a limit $\dms > 3.1$~ps$^{-1}$ 
at 95\% confidence level.
The region 5.0~ps$^{-1} < \dms < 7.6$~ps$^{-1}$ is also excluded.
The dotted curve
shows the result of performing the full minimisation at 
the global minimum,
$\dms=3.6$~ps$^{-1}$,
using the default resolution functions,
and then scanning through $\dms$ while keeping 
all other parameters fixed.
The fitted value of $f_{\mathrm s}$, of particular relevance
for $\dms$ measurements, was $(11.5 \pm 1.9) \%$.

A second technique for
studying $\dms$, known as the amplitude method~\cite{moser}, was also used.
The method consists of replacing the $\cos \dms t'$ term in the
physics function with $A\cos \dms t'$, where $A$ is known as the
amplitude.
The parameter $A$ may be fitted at each value of $\dms$, and if
the value is smaller than unity then the value of $\dms$ may be
excluded with a confidence that depends only on the value and
uncertainty of $A$.
The advantage of such a technique over an exclusion based on 
$\Delta \log \cal L$ is that
$A$ is measured with errors
that are approximately Gaussian.
This makes it easy to combine results and compare sensitivities
from different analyses or experiments.
The calculation of a limit is also relatively straightforward.

\begin{figure}[htbp]
\centering
\epsfxsize=17cm
\begin{center}
    \leavevmode
    \epsffile[20 151 525 674]{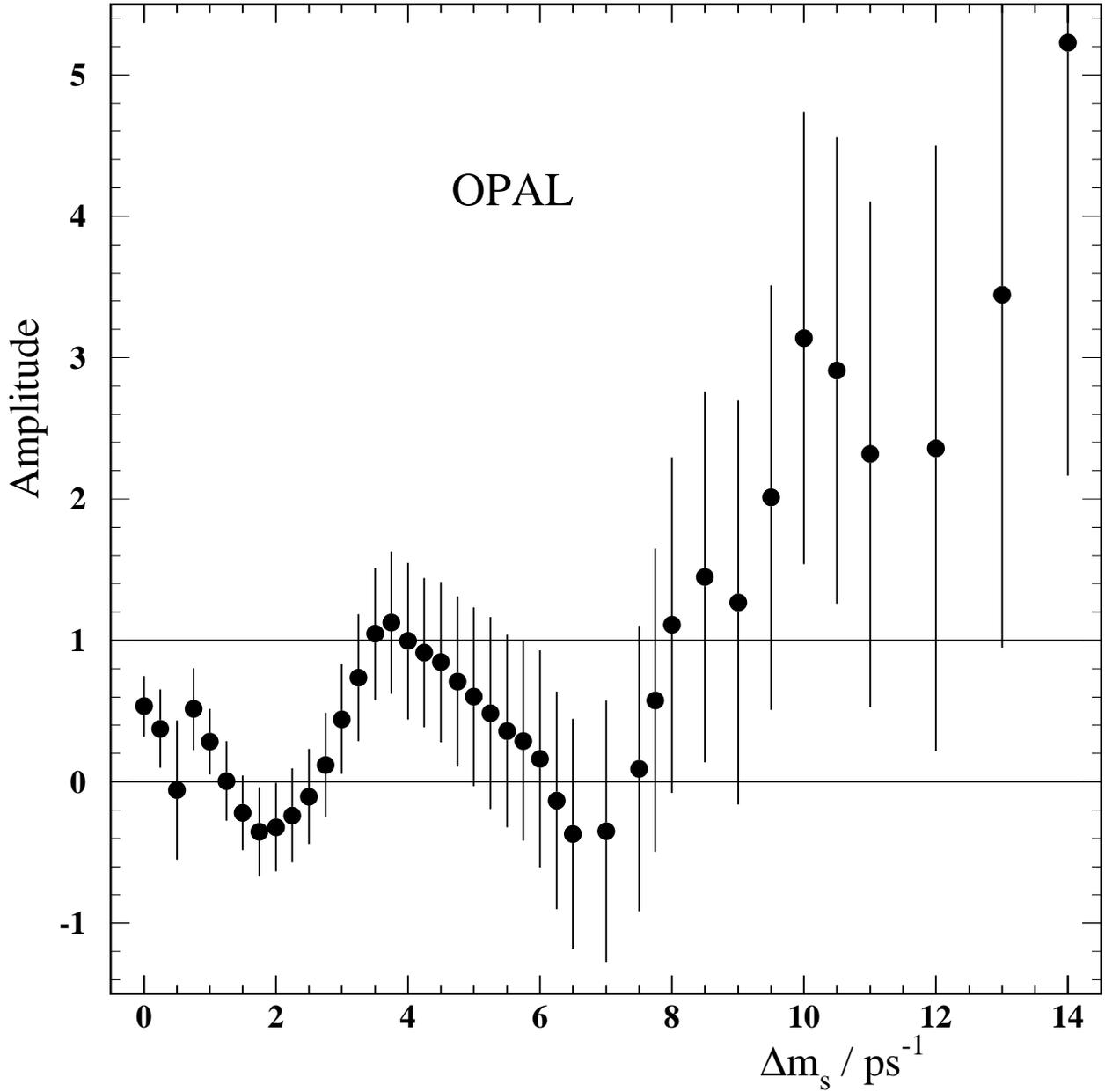}
\end{center}
\vspace{-5 mm}
\caption{The result of the amplitude fit.
The points show the fitted value of the amplitude, $A$, as a function of
$\dms$, and the error bars include all systematic errors.
}
\label{fig:amp}
\end{figure}

For this analysis, $A$ was simply added as an extra parameter to the
fit.
All systematic uncertainties were handled in the same way as for the
$\dmd$ result.
The fitted value of 
$A$ as a function of $\dms$ is shown in
Figure~\ref{fig:amp}.
To facilitate combination with other results,
the central values and a breakdown of the error contributions in steps
of 1~ps$^{-1}$ are documented in the Appendix. 
The results are qualitatively in agreement with the $\Delta \log \cal L$
method.
There is a possible signal near $\dms=4$~ps$^{-1}$,
as seen also in the $\Delta \log \cal L$ plot,
where $A$ is
close to 1 and significantly separated from 0.
Previous measurements~\cite{dsl,Wu} exclude this region
with a high confidence, so that this excursion is more likely to be
a statistical fluctuation.
To determine exclusion regions at 95\% confidence level, at a given
value of $\dms$
we represent
the measured value of $A$ as a Gaussian distribution
function $G(A-\mu,\sigma_A)$, 
where $\mu$ is the central value and $\sigma_A$ is the
measurement error.
Two alternative methods are then considered to determine 
whether the value of $\dms$ is excluded:
\begin{description}
\item[a)]
values are excluded
where the probability of measuring an amplitude lower than
that observed is less than 5\%, if that value of $\dms$ was the correct
one, i.e. 
\begin{equation}\int_1^{\infty} G(A-\mu,\sigma_A) {\mathrm d}A < 0.05 \; ,
\end{equation} or
\item[b)]
the same definition, but limited to the positive region, i.e.
\begin{equation} 
\frac{\int_1^{\infty} G(A-\mu,\sigma_A) {\mathrm d}A}
        {\int_0^{\infty} G(A-\mu,\sigma_A) {\mathrm d}A} < 0.05 \; .
\end{equation}
\end{description}
The first definition gives 
a true 95\% confidence level, in the sense that there
is a 5\% probability to exclude the true value.
However, it is not protected
against setting limits well beyond the experimental sensitivity.
The second definition makes use of the fact that the predicted value
of $A$ lies between 0 and 1, regardless of the value of $\dms$.
It is automatically protected against setting limits beyond the
sensitivity.
For a true value of $\dms$ well beyond the sensitivity, method a)
would exclude the true value in 5\% of the experiments, while for method
b) this percentage would tend towards zero.
%

For method a), excluded regions are defined by 
$A + 1.645\sigma_A < 1$, where $\sigma_A$ represents the total error
on $A$.
This gives the result $\dms > 2.9$~ps$^{-1}$ at 95\% confidence level.
The small region $6.4$~ps$^{-1} < \dms < 6.7$~ps$^{-1}$ is also excluded.
For method b), we find a limit $\dms > 2.9$~ps$^{-1}$,
and higher $\dms$ regions are not excluded.
The difference between the two lower limits is 0.03~ps$^{-1}$.
Both results are similar to those obtained from the 
$\Delta \log \cal L$ technique.

We define a measure of the sensitivity of this experiment
as the value of $\dms$ that would be excluded by method a)
if $A$ were measured to be 0, i.e. where $1.645\sigma_A = 1$.
The sensitivity is $4.8$~ps$^{-1}$. 

\subsection{Data tests}
\begin{figure}[htbp]
\centering
\epsfxsize=17cm
\begin{center}
    \leavevmode
    \epsffile[42 395 503 652]{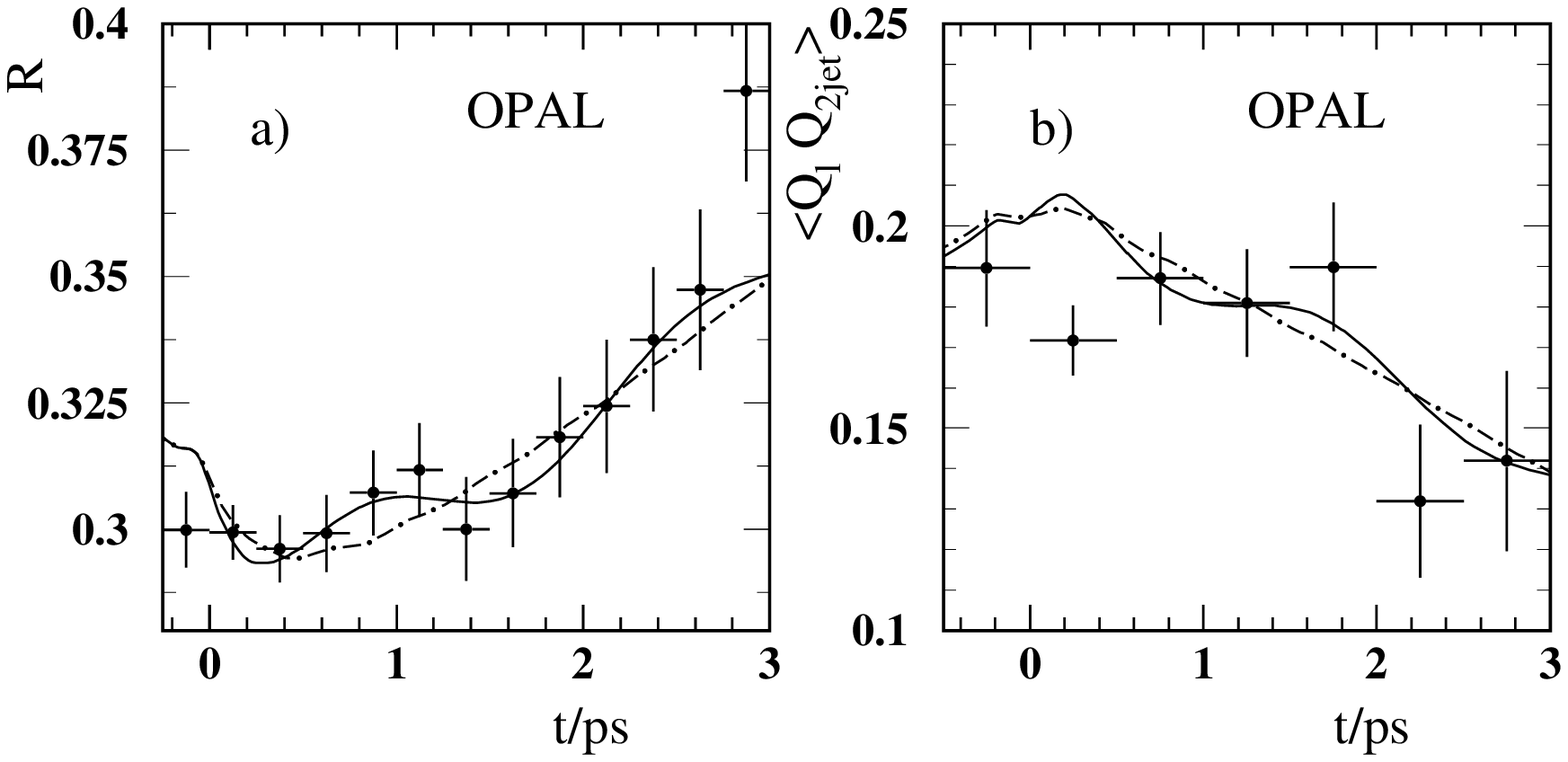}
\end{center}
\vspace{-5 mm}
\caption{ 
a) The ratio $R$ versus $t$
for data with $| Q_{2jet} | > 2$, restricted to the time
window $-0.25$~ps~$< t <$~3~ps.
b) The value of
$\langle Q_{\ell} Q_{2jet} \rangle$
versus $t$ for data with
$| Q_{2jet} | < 2$, for the time window $-0.5$~ps~$< t <$~3~ps.
The solid line shows the result of the fit when $\dms = 3.6$~ps$^{-1}$,
and the dashed-dotted line corresponds to $\dms=15$~ps$^{-1}$. 
}
\label{fig:t2}
\end{figure}

The plots of $R$ (the fraction of events with 
$Q_{\ell} Q_{2jet}< 0$)
versus $t$ and  
$\langle Q_{\ell} Q_{2jet} \rangle$
versus $t$ shown in Figure~\ref{fig:t} are repeated in 
Figure~\ref{fig:t2}, but restricted to the time windows
$-0.25$~ps~$< t <$~3~ps and $-0.5$~ps~$< t <$~3~ps respectively.
Also shown in these plots are the fitted curves assuming
$\dms=3.6$~ps$^{-1}$ and $\dms=15$~ps$^{-1}$ using the nominal
tracking resolution.
%
Calculating the $\chi^2$ summed over both plots 
(over the full time range) yields values of 44.8 for 
$\dms=3.6$~ps$^{-1}$ and 48.0 for $\dms=15$~ps$^{-1}$ for 
39 bins in each case.
The difference in $\log \cal L$ between these two values of $\dms$
is 2.5 (again using the nominal tracking resolution),
showing a behaviour similar to that of
the calculated $\chi^2$ values.
This test suggests that the global minimum at $\dms=3.6$~ps$^{-1}$
is not an artefact of the fitting procedure, but is
rather a property of the data.

To test the stability of the results,
the $\dmd$ fit was repeated using a tighter selection:
the $\nn$ cut was changed from 0.7 to 0.85.
The fit result was $0.438 \pm 0.039$~ps$^{-1}$, 
consistent with the previous result of $0.444 \pm 0.034$~ps$^{-1}$.


\subsection{Monte Carlo tests}

The multiparameter constrained fit described above 
was performed on a Monte Carlo sample based on four million hadronic events
generated with $\dmd=0.438$~ps$^{-1}$ and
an infinite value of $\dms$, with $\dms$ fixed to a
large value
in the fit. 
The fitted value for $\dmd$
was $0.430 \pm 0.030$~ps$^{-1}$.
The fitted values for three of the parameters, 
$\langle \tau_{\mathrm b} \rangle$, $f_{\mathrm uds}$
and $f_{\mathrm{\bcas}}$, were more 
than 2$\sigma$ from their true values.
If the uncertainty due to the resolution functions
and their parametrisation is taken into
account, then all values of all parameters are
consistent with their true values.
A possible bias associated 
with these parameters was
investigated in the data by repeating the fit with
all three of these parameters
fixed to their nominal values. 
The result for $\dmd$ was shifted by $+0.003$~ps$^{-1}$.
The effect on the fitted amplitude for $\Bs$ oscillations
at $\dms=3.75$~ps$^{-1}$ was a shift of 0.18, 
which is easily covered by the total error of 0.50.
The possible bias does not cause a significant problem
for the results, and is addressed by resolution function
uncertainties.

\begin{figure}[htbp]
\centering
\epsfxsize=17cm
\begin{center}
    \leavevmode
    \epsffile[22 147 530 670]{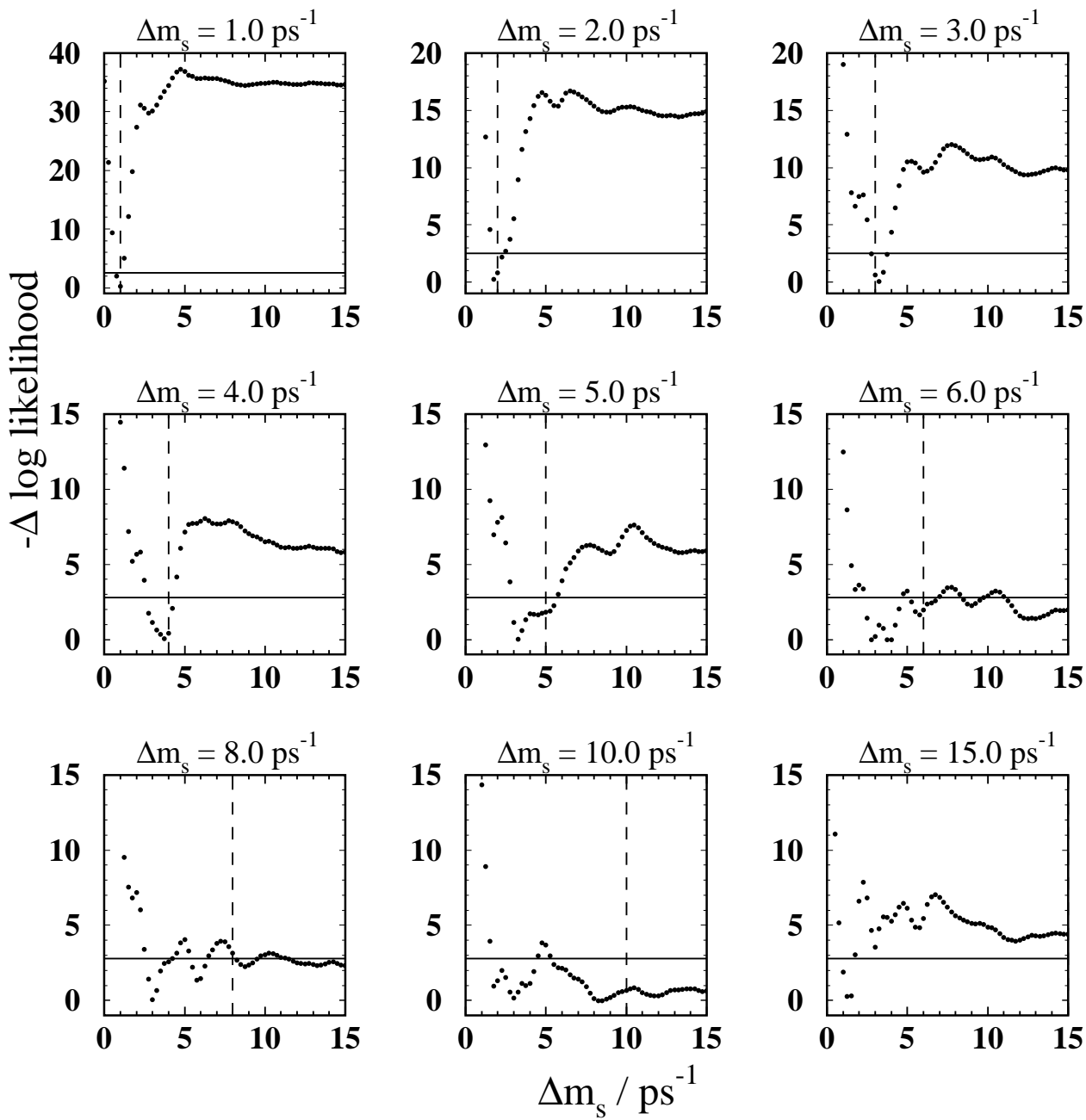}
\end{center}
\vspace{-5 mm}
\caption{ 
The value of
$- \Delta \log \cal L$, relative
to the maximum value of $\log \cal L$, as a function  
of $\dms$ for data simulated using nine different
$\dms$ values, indicated by the dashed line in each case.
All other parameters in the likelihood calculation
were fixed.
The Monte Carlo event statistics are equivalent to 4 million hadronic $\Z$ decays.
Note that 
the different values of $\dms$ are not statistically independent.
Horizontal lines are drawn on each plot at 
$- \Delta \log {\cal L} = 2.8$, corresponding roughly
to a 95\% confidence level.
}
\label{fig:mc}
\end{figure}

To test the sensitivity to $\dms$, the single sample of
four million simulated events
was used to simulate
event samples with $\dmd =0.45$~ps$^{-1}$ and a range
of values for $\dms$: 1~ps$^{-1}$, 2~ps$^{-1}$, 3~ps$^{-1}$, 
4~ps$^{-1}$,
5~ps$^{-1}$, 6~ps$^{-1}$, 8~ps$^{-1}$, 10~ps$^{-1}$ and 15~ps$^{-1}$.
The values of the fit parameters other
than $\dms$ were fixed at their known values in the simulation,
and $\log \cal L$ calculated as a function of $\dms$ for each simulated
event sample.
The resulting plots of $- \Delta \log \cal L$ versus $\dms$
(relative to the point of maximum likelihood) 
for simulated data are shown in 
Figure~\ref{fig:mc}.
%
It can be seen that $\Delta \log \cal L$, calculated at the 
generated value,
is in most cases smaller than 2.8, which corresponds roughly
to a 95\% confidence level.
The exceptions are the samples generated with $\dms=8$~ps$^{-1}$,
where the generated value is just excluded,
and $\dms=15$~ps$^{-1}$, but note that the samples are statistically
correlated with each other.
It is interesting to note that in the case of 
the sample generated with $\dms=15$~ps$^{-1}$,
an apparent signal would have been seen at a low value of $\dms$.
Such cases are properly included in the fast Monte Carlo simulations.

\begin{figure}[htbp]
\centering
\epsfxsize=17cm
\begin{center}
    \leavevmode
    \epsffile[22 147 530 670]{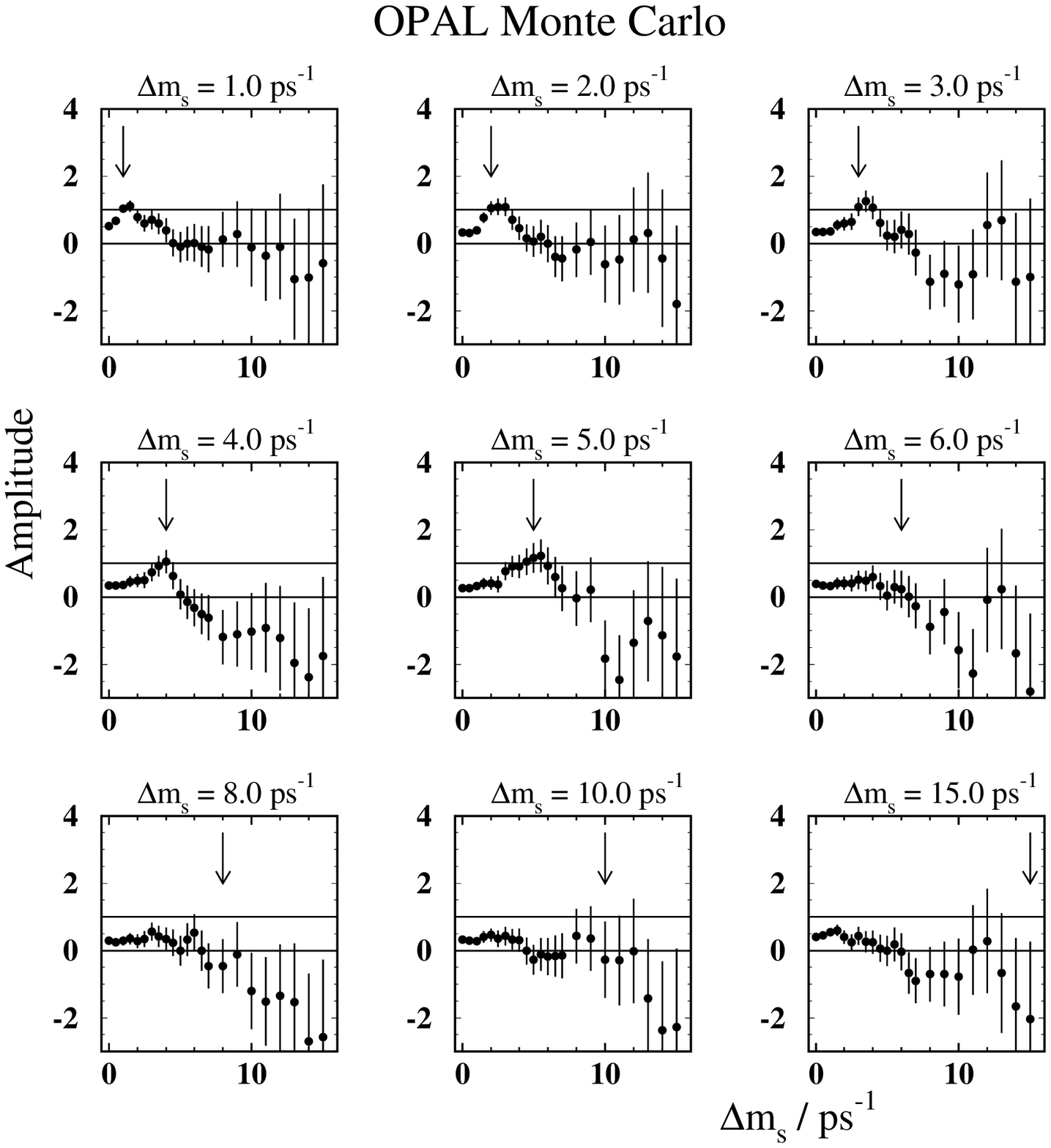}
\end{center}
\vspace{-5 mm}
\caption{
The fitted value of $A$ as a function of $\dms$
for data simulated using nine different $\dms$ values,
indicated by the arrow in each case.
All parameters other than $A$ were fixed in the fit.
The Monte Carlo event statistics are equivalent to 4 million hadronic $\Z$ decays,
but the nine samples are statistically correlated.
}
\label{fig:mcamp}
\end{figure}

Using the same simulated samples, the equivalent test was 
performed for the amplitude method.
All fit parameters were fixed with the exception of $A$, 
which was fitted as a function of $\dms$, for each sample.
The results are shown in Figure~\ref{fig:mcamp}.
For the sample generated with $\dms=8$~ps$^{-1}$, 
the generated value of $\dms$ would be excluded
by method a), but not by method b).
In none of the other samples would the generated
$\dms$ value be excluded by either method.  

\section{Test of CP(T) conservation}

The multiparameter fit was repeated with two extra free parameters,
$\reps$ and $\idel$, with the physics function modified as described
in section~5.
The results were first tested on four Monte Carlo samples simulated
with and without CP and CPT violation.
The four Monte Carlo samples are statistically correlated and are
each equivalent to four million hadronic events.
The results of this test, in which $\dms$ was fixed to 
a large value (and $A$ fixed to 1), 
but all other parameters were free to vary, are given 
in Table~\ref{tab:mccpt}.
\begin{table}[htbp]
\begin{center}
\begin{tabular}{|c|c|c|c|} \hline
Generated $\reps$ & Generated $\idel$ & Fitted $\reps$ & Fitted
$\idel$ \\ \hline
0 & 0   & $-0.014 \pm 0.009$ & $0.005 \pm 0.014$ \\
0 & 0.1 & $-0.015 \pm 0.010$ & $0.096 \pm 0.015$ \\
0.1 & 0 & $0.077~^{+0.012}_{-0.011}$ & $-0.002~^{+0.015}_{-0.014}$ \\
0.2 & 0 & $0.177~^{+0.018}_{-0.016}$ & $-0.013 \pm 0.016$ \\ \hline
\end{tabular}
\caption{Results of fits for $\reps$ and $\idel$ using
simulated data.
There is a strong statistical correlation between the different
samples.} 
\label{tab:mccpt}
\end{center}
\end{table}

\begin{figure}[htbp]
\centering
\epsfxsize=17cm
\begin{center}
    \leavevmode
    \epsffile[21 157 531 649]{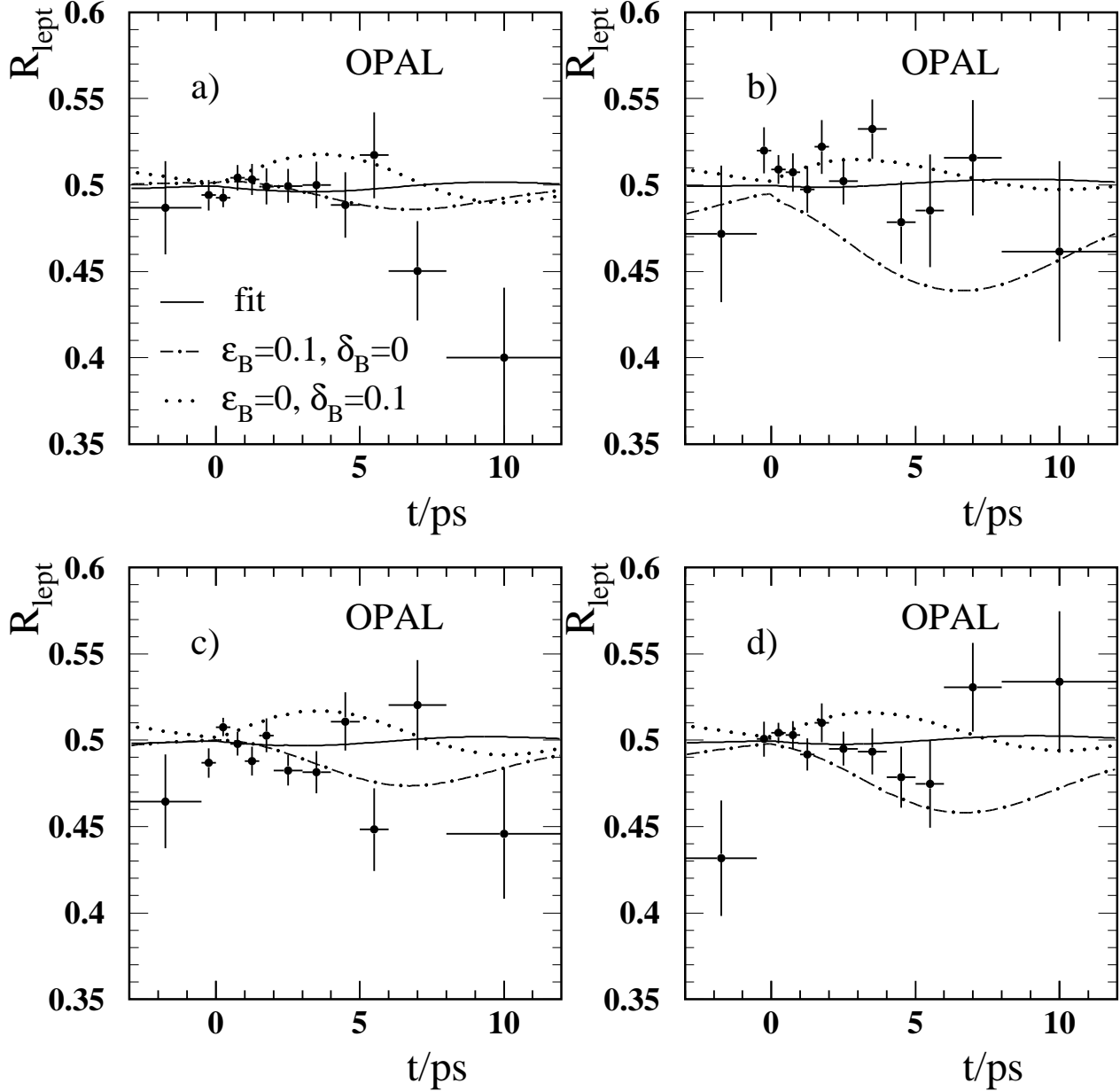}
\end{center}
\vspace{-5 mm}
\caption{ 
The distribution of $R_{\mathrm lept}$ versus reconstructed time for
the data, with the result of the CPT fit superimposed.
The data is divided into four categories :
a) $Q_{\ell} Q_{2jet} > 0$ and $| Q_{2jet} | > 2$,
b) $Q_{\ell} Q_{2jet} < 0$ and $| Q_{2jet} | > 2$,
c) $Q_{\ell} Q_{2jet} > 0$ and $| Q_{2jet} | < 2$,
d) $Q_{\ell} Q_{2jet} < 0$ and $| Q_{2jet} | < 2$.
Also shown are the predicted curves for $\reps=0.1$, $\idel=0$
and for $\reps=0$, $\idel=0.1$.
}
\label{fig:cpt}
\end{figure}

The result of the fit to the data, with $\dms$ fixed to 15~ps$^{-1}$,
was
\begin{eqnarray}
\reps & = & -0.006 \pm 0.010 \nonumber \\
\idel & = & -0.020 \pm 0.017 \; .\nonumber
\end{eqnarray}
The distribution of $R_{\mathrm lept}$,
the fraction of leptons that are negatively charged,
versus reconstructed time
is shown in Figure~\ref{fig:cpt} for the data,
 with the fit result
superimposed.
In this figure, the data is divided into four categories according
to the sign of $Q_{\ell} Q_{2jet}$ and whether $| Q_{2jet} |$
is smaller or larger than 2.
Note that, following equation 10, the data with 
$Q_{\ell} Q_{2jet} < 0$ (enhanced in mixed events) are more
sensitive to $\reps$, while the data with 
$Q_{\ell} Q_{2jet} > 0$ are more sensitive to $\idel$.
For comparison, the predicted curves assuming $\reps=0.1$, $\idel=0$
and also $\reps=0$, $\idel=0.1$ are included in the figure.
These curves were obtained keeping all other parameters fixed
at their fitted values.

As in the case of the $\dmd$ analysis, the errors quoted contain
both statistical and systematic components.
The systematic errors are shown in Table~\ref{tab:cpt}, including
the errors from the resolution function, calculated as in the
$\dmd$ analysis.
The uncertainty due to the weighting function $w( |Q_{2jet}| )$ is also
indicated, estimated by shifting the jet charge by 0.06 
for Monte Carlo simulated with $\chi=0.5$ in the opposite jet.
Such a shift causes a 10\% change in $w( |Q_{2jet}| )$.
\begin{table}[htbp]
\begin{center}
\begin{tabular}{|c|c|c||c|} \hline
Origin & Error on $\reps$ & Error on $\idel$ & Error on $\reps$, $\idel=0$\\ \hline
$f_{\mathrm s}$       & $\pm 0.000 $ & $\pm 0.000 $ & $\pm 0.000$\\
$f_{\mathrm{baryon}}$ & $\pm 0.002 $ & $\pm 0.002 $ & $\pm 0.000$\\
$f_{\bcas}$           & $\pm 0.000 $ & $\pm 0.000 $ & $\pm 0.000$\\
$f_{\mathrm c}$       & $\pm 0.000 $ & $\pm 0.000 $ & $\pm 0.000$ \\
$f_{\mathrm{uds}}$    & $\pm 0.000 $ & $\mp 0.001 $ & $\pm 0.000$ \\
 & & &\\
$\langle \tau_{\mathrm b} \rangle $ & $\mp 0.001$ & $\mp 0.001 $ & $\pm 0.000$ \\
$\tau^+ / \taud $           & $\pm 0.000$ & $\pm 0.001$ & $\pm 0.000$\\
$\tau_{\mathrm s} / \taud $ & $\mp 0.001$ & $\pm 0.000$ & $\pm 0.000$ \\
$\tau_{\Lb} / \taud $       & $\mp 0.001$ & $\mp 0.001$ & $\pm 0.000$ \\
 & & &\\
$\delta Q_{\mathrm b}$     & $\pm 0.000$ & $\pm 0.000$ & $\pm 0.000$ \\
$\delta Q_{\mathrm{mix}}$  & $\pm 0.001$ & $\pm 0.001$ & $\pm 0.000$ \\
$\delta Q_{\Bp}$           & $\pm 0.001$ & $\pm 0.002$ & $\pm 0.000$ \\
$\delta Q_{\mathrm{udsc}}$ & $\pm 0.001$ & $\pm 0.001$ & $\pm 0.000$ \\
$f_{\mathrm D^{**}}$       & $\pm 0.000$ & $\pm 0.000$ & $\pm 0.000$ \\ 
 & & &\\
Tracking resolution        & $\pm 0.001$ & $\pm 0.001$ & $\pm 0.000$ \\
b fragmentation            & $\pm 0.001$ & $\pm 0.002$ & $\pm 0.000$ \\
Resolution effect of $\mathrm D^{**}$ & $\pm 0.000$ & $\pm 0.000$ & $\pm 0.000$ \\
$\dms$ variation           & $\pm 0.000$ & $\pm 0.000$ & $\pm 0.000$ \\ 
$w( |Q_{2jet}| )$          & $\pm 0.001$ & $\pm 0.000$ & $\pm 0.000$\\
 & & &\\
Jet charge asymmetry    & $\pm 0.004$ & $\pm 0.003$ & $\pm 0.003$ \\
Lepton charge asymmetry & $\pm 0.001$ & $\pm 0.001$ & $\pm 0.001$ \\
Momentum asymmetry      & $\pm 0.001$ & $\pm 0.001$ & $\pm 0.001$ \\
Lepton background       & $\pm 0.002$ & $\pm 0.003$ & $\pm 0.001$ \\
$\cos \theta$ asymmetry & $\pm 0.000$ & $\pm 0.000$ & $\pm 0.000$\\ \hline
Total                    & $\pm 0.006$ & $\pm 0.006$ & $\pm 0.003$\\ \hline
\end{tabular}
\caption{Summary of systematic errors on $\reps$ and $\idel$. 
The fourth column gives the systematic errors on $\reps$ when $\idel$
is fixed to 0.} 
\label{tab:cpt}
\end{center}
\end{table}

In addition to these uncertainties, extra sources of error
that affect only the CP(T) parameters had to be considered.
Firstly, the mean jet charge for produced b quarks is not
exactly opposite that for produced $\mathrm \bar{b}$ quarks.
This is caused by a slight asymmetry in the jet chamber
in the efficiency to detect low momentum positively and
negatively charged tracks.
The asymmetry arises from the geometry of the detector and
is therefore simulated in the Monte Carlo.
The mean jet charge, averaged over b and $\mathrm \bar{b}$ quarks,
is found to be $+0.020 \pm 0.006$, where the error is statistical
only.
This result was checked by studying the charge
asymmetry for tracks in hadronic events
for both the data and the Monte Carlo.
The mean jet charge could then be calculated
by weighting the charge asymmetry appropriately
as a function of momentum, according to the
contribution to the jet charge.
Consistent results were obtained whether the weighting was determined
from data or Monte Carlo.
The resulting mean jet charge values were 0.008
for data and 0.013 for Monte Carlo.
This calculated Monte Carlo result is consistent
with the value directly observed in the Monte Carlo.
The final value for the data is determined
by scaling 0.008 by $0.020/0.013$, assigning
an error of 0.006 due to Monte Carlo statistics,
and 0.008 due to the discrepancy seen between 
Monte Carlo and data, to give $0.012 \pm 0.010$.
This tracking asymmetry could also have a minor effect on the
efficiency for selecting leptons (required to have $p >2$~GeV/c).
In this case, the statistics of the Monte Carlo are insufficient to
investigate the effect directly.
The effect was checked by studying inclusive tracks, passing the
same track quality cuts as the leptons, as a function of $p$.   
A possible asymmetry of $6\times 10^{-4}$ was deduced for a 
value of $\langle 1/p \rangle = 0.14$~GeV$^{-1}$, corresponding
to the lepton sample.
This was assigned as a systematic uncertainty.
Another cause of uncertainty would be an
asymmetry in the momentum measurements for positive
and negative tracks, which could be caused by a  slight rotation of the
outside of jet chamber relative to the inside.
This effect was investigated by studying $\Zzero \arr \mumu$ events
in the data. 
A difference in $\langle 1/p \rangle$ of 
$(7 \pm 2)\times 10^{-5}$~GeV$^{-1}$ was
observed between $\mu^-$ and $\mu^+$ particles.
Such a shift could cause a lepton selection asymmetry of 
$5\times 10^{-4}$, assigned as a systematic error.

The results are also sensitive to a charge asymmetry
in the lepton background.
This asymmetry was studied in the data using electron
and muon candidates that just failed the lepton
selection criteria.
The charge asymmetry was found to be $(1.9 \pm 0.7)$\%,
with the largest component due to muon candidates
resulting from non-interacting kaons and kaon punchthrough.
The central values quoted above were corrected for this
effect.

Finally, a $\cos \theta$ asymmetry for lepton identification
in the detector would cause
a charge asymmetry through the $\Zbb$ forward-backward
asymmetry, measured by OPAL to be $(9.06 \pm 0.51 \pm 0.23)\%$ at the
$\Zzero$ peak \cite{afb}.
CP(T) violation would not induce a $\cos \theta$ asymmetry. 
A lepton $\cos \theta$ asymmetry of $(-1.7 \pm 0.3)$\% (more leptons at
$\cos\theta < 0$) was observed in the data.
This can be understood as due to an observed
average shift of about +0.5~cm of the
beam spot along the $z$-axis relative to the centre of the detector.  
The central values of $\reps$ and $\idel$ quoted above were 
adjusted by $-0.002$ and $-0.001$, respectively,
for this asymmetry
and
20\% (representing the relative error on the $\cos \theta$ asymmetry)
of the shifts taken as systematic error.

Splitting the fit errors into their statistical and systematic
components and including the extra systematic errors described above,
the final results are :
\begin{eqnarray}
\reps & = & -0.006 \pm 0.010 \pm 0.006 \nonumber \\
\idel & = & -0.020 \pm 0.016 \pm 0.006 \; .\nonumber
\end{eqnarray}
These results neglect CP violation in the $\Bs$ system.
The sensitivity of the results to $\repsbs$, 
where $\epsilon_{\Bs}$ is $\epsilon$ for the $\Bs$ system, was gauged
by repeating the fit assuming a large CP asymmetry,
$\repsbs=-0.05$.
The value of $\reps$ was shifted by +0.015, while $\idel$
was shifted by +0.006.
Note that the influence of $\idelbs \neq 0$ on the results
is negligible, because
the effect is smeared out by the rapid oscillations (see equation 10).

CPT is usually assumed to be a good symmetry, i.e. 
$\delta_{\mathrm b}= 0$. 
In the results presented above, there is a correlation of order
+30\% between the $\reps$ and $\idel$ results, so we
can obtain more precise $\reps$ results if $\idel$ is set to 0.
The fit was repeated with $\idel =0$ to give
\[
\reps = 0.002 \pm 0.007 \; ,
\]
where the fit contains both statistical and systematic errors.
The systematic error calculation was performed as described 
above, and the errors are included in Table~\ref{tab:cpt}.
The final result is 
\[
\reps = 0.002 \pm 0.007 \pm 0.003 \; .
\]
In this case, the effect of $\repsbs=-0.05$ causes a shift of
+0.013 in $\reps$.
 
\section{Conclusions}

A sample of inclusive lepton events was used to study $\Bd$ and $\Bs$
oscillations.
An estimate of the decay proper time 
of the inferred b hadron was reconstructed
for each lepton candidate, and a jet charge technique was used
to tag the produced B flavour.
We measure 
\[ 
\dmd = 0.444 \pm 0.029~^{+0.020}_{-0.017}~\mathrm{ps}^{-1} \; .
\]
This result is consistent with 
previous measurements
\cite{hideto,shlomit,deltamd},
and represents the most precise result for a single technique.
Taking account of the common systematic errors,
we combine this measurement with previous OPAL
results~\cite{hideto,shlomit}
to give
\[ \dmd = 0.475 ^{+0.022}_{-0.023}~^{+0.018}_{-0.016}~\mathrm{ps}^{-1} . \]
The small statistical correlations between the results
have a negligible effect.

By studying the behaviour of $\log \cal L$ as a function of $\dms$,
we are able to exclude the regions
$ \dms < 3.1~{\mathrm{ps}}^{-1}$ and
$5.0~{\mathrm{ps}}^{-1} < \dms <7.6~{\mathrm{ps}}^{-1}$
at 95\% confidence level.
Using an amplitude method instead, the lower limit of 3.1~ps$^{-1}$
is slightly weakened to 2.9~ps$^{-1}$, and the region 
5.0~ps$^{-1} < \dms < 7.6$~ps$^{-1}$ 
is either only partially excluded or not excluded 
at all, depending on the details of the confidence level calculation.
The lower limit that would be obtained using the amplitude method,
were the amplitude measured to be zero, is 4.8~ps$^{-1}$,
a measure of the experimental sensitivity.
This result is consistent with previous
results~\cite{hideto,aleph,Wu,dsl}, of which 
the most constraining~\cite{dsl} quotes a lower limit of 6.6~ps$^{-1}$
at 95\% confidence level.

By studying the charge symmetry of the $\Bd$ mixing structure,
we are able to constrain possible CP and CPT violating
effects.
We measure simultaneously the indirect CP violation parameter
\[ \reps = -0.006 \pm 0.010 \pm 0.006
\]
and the indirect CPT violation parameter
\[ \idel = -0.020 \pm 0.016 \pm 0.006 \;\; .
\]
If alternatively we invoke CPT symmetry, then we obtain
\[
\reps = 0.002 \pm 0.007 \pm 0.003 \; .
\]
The $\reps$ measurement, whether or not CPT symmetry is invoked,
is consistent with, but more precise than,
previous measurements from CLEO~\cite{cl_eps}
and CDF~\cite{cdf_eps}.
The $\idel$ result represents the first measurement of this
quantity in the B system.

\section*{Appendix: Amplitude results}
Section~6.2 describes the amplitude results for the $\dms$ study.
We detail in Table~\ref{tab:amp} the central values of the
amplitude and the breakdown of the error contributions
in steps of 1~ps$^{-1}$.
This information is essential for a correct combination 
of the results from this paper with other analyses.
The systematic uncertainties for intermediate points
may be estimated by interpolation.
\begin{table}[htbp]
\begin{center}
{\small
\setlength{\doublerulesep}{0.5cm}
\begin{tabular}{|c|c|c|c|c|c|c|c|c|} \hline
$\dms$ (ps$^{-1}$) & 0 & 1 & 2 & 3 & 4 & 5 & 6 & 7 \\ \hline
$A$ &0.53 &0.29 & $-0.32$ &0.44 &0.99 &0.60 &0.16 & $-0.35$ \\
$\sigma_A^{\mathrm{stat}}$ & $\pm 0.18$& $\pm 0.19$& $\pm 0.29$& $\pm
0.35$& $\pm 0.46$& $\pm 0.59$& $\pm 0.75$& $\pm 0.91$ \\ \hline
$f_{\mathrm s}$ & $\pm 0.09$ & $\pm 0.11$ & $\pm 0.01$ & $\mp 0.08$ &
$\mp 0.23$ & $\mp 0.10$ & $\mp 0.08$ & $\pm 0.00$ \\
$f_{\mathrm{baryon}}$ & $\pm 0.01$& $\pm 0.01$& $\mp 0.04$& $\pm 0.02$& $\mp 0.02$& $\pm 0.00$& $\pm 0.02$& $\pm 0.03$ \\
$f_{\bcas}$ & $\pm 0.04$& $\pm 0.00$& $\mp 0.02$& $\mp 0.02$& $\mp 0.01$& $\mp 0.02$& $\mp 0.01$& $\pm 0.00$ \\
$f_{\Bs \arr {\mathrm c} \arr \ell}$ & $\pm 0.01$& $\pm 0.00$& $\pm 0.00$& $\pm 0.01$& $\pm 0.04$& $\pm 0.01$& $\pm 0.02$& $\pm 0.03$ \\
$f_{\mathrm c}$ & $\pm 0.00$& $\mp 0.01$& $\mp 0.01$& $\pm 0.00$& $\mp 0.01$& $\mp 0.01$& $\mp 0.01$& $\pm 0.00$ \\
$f_{\mathrm{uds}}$ & $\pm 0.00$& $\pm 0.01$& $\pm 0.01$& $\pm 0.00$& $\pm 0.01$& $\pm 0.01$& $\pm 0.01$& $\pm 0.00$ \\
$\langle \tau_{\mathrm b} \rangle $ & $\pm 0.00$& $\pm 0.01$& $\pm 0.00$& $\pm 0.00$& $\mp 0.01$& $\pm 0.00$& $\pm 0.00$& $\mp 0.01$ \\
$\tau^+ / \taud $ & $\mp 0.03$& $\mp 0.03$& $\pm 0.01$& $\mp 0.01$& $\mp 0.01$& $\mp 0.01$& $\mp 0.02$& $\pm 0.00$ \\
$\tau_{\mathrm s} / \taud $ & $\pm 0.02$& $\pm 0.02$& $\pm 0.00$& $\mp 0.03$& $\mp 0.01$& $\mp 0.02$& $\mp 0.01$& $\mp 0.03$ \\
$\tau_{\Lb} / \taud $ & $\mp 0.01$& $\pm 0.01$& $\mp 0.02$& $\pm 0.00$& $\mp 0.02$& $\pm 0.00$& $\pm 0.00$& $\pm 0.01$ \\
$\delta Q_{\mathrm b}$ & $\mp 0.06$& $\mp 0.02$& $\pm 0.02$& $\mp 0.02$& $\mp 0.06$& $\pm 0.00$& $\mp 0.01$& $\pm 0.05$ \\
$\delta Q_{\mathrm{mix}}$ & $\mp 0.01$& $\mp 0.01$& $\pm 0.00$& $\mp 0.02$& $\mp 0.08$& $\mp 0.03$& $\mp 0.02$& $\pm 0.01$ \\
$\delta Q_{\Bp}$ & $\mp 0.03$& $\mp 0.02$& $\mp 0.02$& $\pm 0.02$& $\pm 0.04$& $\pm 0.01$& $\pm 0.01$& $\mp 0.04$ \\
$\delta Q_{\mathrm{udsc}}$ & $\mp 0.01$& $\pm 0.00$& $\mp 0.05$& $\mp 0.07$& $\mp 0.09$& $\mp 0.10$& $\mp 0.10$& $\mp 0.11$ \\
$f_{\mathrm D^{**}}$ & $\pm 0.01$& $\pm 0.01$& $\pm 0.00$& $\pm 0.00$& $\pm 0.04$& $\pm 0.01$& $\mp 0.01$& $\mp 0.01$ \\ 
Tracking resolution & $\pm 0.01$& $\pm 0.03$& $\pm 0.08$& $\pm 0.11$& $\pm 0.15$& $\pm 0.14$& $\pm 0.12$& $\pm 0.01$ \\
b fragmentation & $\pm 0.00$& $\pm 0.02$& $\pm 0.02$& $\pm 0.02$& $\pm 0.01$& $\pm 0.00$& $\pm 0.01$& $\pm 0.04$ \\
$\Bs$ resolution function & $\pm 0.01$& $\pm 0.03$& $\pm 0.03$& $\pm 0.01$& $\pm 0.04$& $\pm 0.09$& $\pm 0.01$& $\pm 0.06$ \\ \hline
$\sigma_A^{\mathrm{syst}}$ & $\pm 0.13$& $\pm 0.14$& $\pm 0.12$& $\pm
0.16$& $\pm 0.31$& $\pm 0.23$& $\pm 0.18$& $\pm 0.16$ \\  \hline \hline
$\dms$ (ps$^{-1}$) & 8 & 9 & 10 & 11 & 12 & 13 & 14 & 15 \\ \hline
$A$ & 1.11 & 1.27 & 3.14 & 2.32 & 2.36 & 3.45 & 5.23 & 8.90 \\
$\sigma_A^{\mathrm{stat}}$ & $\pm 1.06$& $\pm 1.25$& $\pm 1.43$& $\pm 1.69$& $\pm 1.97$& $\pm 2.26$& $\pm 2.64$& $\pm 3.01$\\ \hline
$f_{\mathrm s}$ & $\mp 0.18$& $\mp 0.04$& $\mp 0.46$& $\mp 0.09$& $\mp 0.49$& $\pm 0.00$& $\mp 0.71$& $\mp 1.26$ \\
$f_{\mathrm{baryon}}$ & $\mp 0.03$& $\mp 0.01$& $\mp 0.12$& $\pm 0.03$& $\mp 0.07$& $\pm 0.00$& $\mp 0.09$& $\mp 0.41$ \\
$f_{\bcas}$ & $\pm 0.00$& $\mp 0.03$& $\mp 0.04$& $\mp 0.01$& $\mp 0.01$& $\mp 0.04$& $\mp 0.12$& $\mp 0.14$ \\
$f_{\Bs \arr {\mathrm c} \arr \ell}$ & $\pm 0.03$& $\pm 0.02$& $\pm 0.03$& $\pm 0.01$& $\pm 0.01$& $\pm 0.01$& $\pm 0.01$& $\pm 0.05$ \\
$f_{\mathrm c}$ & $\mp 0.03$& $\mp 0.02$& $\mp 0.07$& $\mp 0.04$& $\mp 0.04$& $\mp 0.03$& $\mp 0.13$& $\mp 0.23$ \\
$f_{\mathrm{uds}}$ & $\pm 0.03$& $\pm 0.01$& $\pm 0.07$& $\pm 0.03$& $\pm 0.04$& $\pm 0.03$& $\pm 0.13$& $\pm 0.22$ \\
$\langle \tau_{\mathrm b} \rangle $ &$\mp 0.01$& $\mp 0.01$& $\pm 0.01$& $\pm 0.00$& $\pm 0.02$& $\mp 0.01$& $\pm 0.00$& $\pm 0.01$ \\
$\tau^+ / \taud $ & $\pm 0.00$& $\mp 0.04$& $\pm 0.00$& $\pm 0.01$& $\mp 0.02$& $\mp 0.01$& $\pm 0.02$& $\pm 0.02$\\
$\tau_{\mathrm s} / \taud $ & $\pm 0.00$& $\mp 0.09$& $\pm 0.04$& $\mp 0.03$& $\mp 0.01$& $\mp 0.04$& $\mp 0.10$& $\pm 0.02$ \\
$\tau_{\Lb} / \taud $ & $\mp 0.01$& $\mp 0.02$& $\pm 0.00$& $\pm 0.02$& $\pm 0.04$& $\pm 0.00$& $\pm 0.04$& $\mp 0.07$ \\
$\delta Q_{\mathrm b}$ & $\mp 0.01$& $\pm 0.01$& $\mp 0.07$& $\pm 0.02$& $\mp 0.10$& $\pm 0.02$& $\mp 0.11$& $\mp 0.33$ \\
$\delta Q_{\mathrm{mix}}$ & $\mp 0.06$& $\mp 0.03$& $\mp 0.12$& $\mp 0.03$& $\mp 0.13$& $\mp 0.03$& $\mp 0.18$& $\mp 0.38$ \\
$\delta Q_{\Bp}$ & $\pm 0.01$& $\pm 0.02$& $\pm 0.05$& $\pm 0.04$& $\pm 0.08$& $\pm 0.02$& $\pm 0.03$& $\pm 0.23$ \\
$\delta Q_{\mathrm{udsc}}$ & $\mp 0.12$& $\mp 0.14$& $\mp 0.14$& $\mp 0.17$& $\mp 0.20$& $\mp 0.19$& $\mp 0.21$& $\mp 0.23$\\ 
$f_{\mathrm D^{**}}$ &  $\pm 0.01$& $\pm 0.01$& $\pm 0.00$& $\mp 0.03$& $\mp 0.02$& $\pm 0.02$& $\pm 0.04$& $\pm 0.04$\\ 
Tracking resolution & $\pm 0.02$& $\pm 0.01$& $\pm 0.06$& $\pm 0.06$& $\pm 0.18$& $\pm 0.29$& $\pm 0.45$& $\pm 0.61$ \\
b fragmentation & $\pm 0.38$& $\pm 0.47$& $\pm 0.31$& $\pm 0.10$& $\pm 0.45$& $\pm 0.79$& $\pm 0.61$& $\pm 0.43$ \\
$\Bs$ resolution function & $\pm 0.31$& $\pm 0.45$& $\pm 0.38$& $\pm 0.52$& $\pm 0.55$& $\pm 0.59$& $\pm 1.07$& $\pm 1.56$ \\ \hline
$\sigma_A^{\mathrm{syst}}$ & $\pm 0.55$& $\pm 0.68$& $\pm 0.72$& $\pm 0.58$& $\pm 0.93$& $\pm 1.05$& $\pm 1.54$& $\pm 2.29$ \\ \hline
\end{tabular}
}
\caption{Amplitude results with the breakdown of systematic errors.} 
\label{tab:amp}
\end{center}
\end{table}
%
%
\par
\vspace*{1.cm}
\section*{Acknowledgements}
\noindent
We particularly wish to thank the SL Division for the efficient operation
of the LEP accelerator and for their continuing close cooperation with
our experimental group. We thank our colleagues from CEA, DAPNIA/SPP,
CE-Saclay for their efforts over the years on the time-of-flight and
trigger
systems which we continue to use.  In addition to the support staff at our
own
institutions we are pleased to acknowledge the  \\
Department of Energy, USA, \\
National Science Foundation, USA, \\
Particle Physics and Astronomy Research Council, UK, \\
Natural Sciences and Engineering Research Council, Canada, \\
Israel Science Foundation, administered by the Israel
Academy of Science and Humanities, \\
Minerva Gesellschaft, \\
Benoziyo Center for High Energy Physics,\\
Japanese Ministry of Education, Science and Culture (the
Monbusho) and a grant under the Monbusho International
Science Research Program,\\
German Israeli Bi-national Science Foundation (GIF), \\
Direction des Sciences de la Mati\`ere du Commissariat \`a l'Energie
Atomique, France, \\
Bundesministerium f\"ur Bildung, Wissenschaft,
Forschung und Technologie, Germany, \\
National Research Council of Canada, \\
Hungarian Foundation for Scientific Research, OTKA T-016660,
T023793 and OTKA F-023259.\\

\end{document}